\documentclass[journal=jacsat,manuscript=article]{achemso}

\usepackage{chemformula}
\usepackage[T1]{fontenc} 
\usepackage[version=3]{mhchem}
\usepackage{amsmath}
\usepackage{amssymb}
\usepackage{hyperref}
\usepackage{booktabs}
\usepackage{graphicx}
\usepackage{pdfpages}

\SectionNumbersOn

\newcommand{\rlg}[1]{#1}

\newcommand{\af}[1]{#1}
\newcommand{\phb}[1]{#1}

\author{M. Gaillard}
\author{A. Faure}
\author{P. Hily-Blant}
\affiliation{Université Grenoble Alpes, CNRS, IPAG, F-38000 Grenoble, France}
\author{R. Le Gal}
\email{romane.le-gal@univ-grenoble-alpes.fr}
\affiliation{Université Grenoble Alpes, CNRS, IPAG, F-38000 Grenoble, France}
\alsoaffiliation{IRAM, 300 rue de la piscine, F-38406 Saint-Martin d’H\`{e}res, France}
\author{S. Lee}
\affiliation{Korea Astronomy and Space Science Institute, 776 Daedeok-daero, Yuseong-gu, Daejeon 34055, Republic of Korea}
\author{H. Nomura}
\affiliation{National Astronomical Observatory of Japan, 2-21-1 Osawa, Mitaka, Tokyo 181-8588, Japan}
\alsoaffiliation{Department of Astronomical Science, The Graduate University for
Advanced Studies, SOKENDAI, 2-21-1 Osawa, Mitaka, Tokyo, 181-8588,
Japan}
\author{K. Furuya}
\affiliation{RIKEN Pioneering Research Institute, 2-1 Hirosawa, Wako-shi, Saitama,
351-0198, Japan}
\alsoaffiliation{Department of Astronomy, Graduate School of Science, The University
of Tokyo, Tokyo, 113-0033, Japan}

\title{Ortho-Para Chemistry of H$_2$CO in the Protoplanetary Disk TW~Hya}

\begin{document}
\newcommand \lp   {\ensuremath{\left(}}
\newcommand \rp   {\ensuremath{\right)}}
\newcommand \lc   {\ensuremath{\left[}}
\newcommand \rc   {\ensuremath{\right]}}
\newcommand \mc[2]{\multicolumn{#1}{c}{#2}}
\newcommand{\pfrac}[2]{\ensuremath{\lp\frac{#1}{#2}}\rp}
\newcommand{\python}{\texttt{python}}
\newcommand{\fortran}{\texttt{fortran 90}}
\newcommand{\ugan}{\texttt{UGAN}}
\newcommand{\collapse}{\texttt{COLLAPSE}}
\newcommand{\tabnote}{\footnotesize\flushleft\hspace{2em}\textsc{Note}---}
\newcommand{\tabnotes}{\footnotesize\flushleft\hspace{1em}\textsc{Notes}---}

\newcommand{\kb}{\ensuremath{k_\text{B}}}
\newcommand{\amu}{\ensuremath{m_\text{a}}}
\newcommand{\stefan}{\ensuremath{\sigma_\text{SB}}}

\newcommand{\erf}{\ensuremath{\text{erf}}}
\newcommand{\erfc}{\ensuremath{\text{erfc}}}
\newcommand{\half}{\ensuremath{\frac{1}{2}}}
\newcommand{\tdix}[1]{\ensuremath{{\,\times 10^{#1}}}}

\newcommand \unit[2]{\ensuremath{\,\text{#1}^{#2}}}
\newcommand \cc{\unit{cm}{-2}}
\newcommand \ccc{\unit{cm}{-3}}
\newcommand{\per}[1]{\ensuremath{\,\text{#1}^{-1}}}
\newcommand{\pper}[1]{\ensuremath{\,\text{#1}^{-2}}}
\newcommand \pers{\per{s}}
\newcommand \cccs{\unit{cm}{3}\,\pers}
\newcommand{\au}{\ensuremath{\text{au}}}
\newcommand{\mstar}{\ensuremath{M_{\star}}}
\newcommand{\msol}{\ensuremath{M_\odot}}

\newcommand{\tkin}{\ensuremath{T_\text{kin}}}
\newcommand{\NH}{\ensuremath{N_\text{H}}}
\newcommand{\nh}{\ensuremath{n_\text{H}}}
\newcommand{\lsol}{\ensuremath{L_\odot}}
\newcommand{\av}{\ensuremath{A_V}}
\newcommand{\depl}{\ensuremath{\delta}}
\newcommand{\depc}{\ensuremath{\delta_\text{C}}}
\newcommand{\depn}{\ensuremath{\delta_\text{N}}}
\newcommand{\tfreez}{\ensuremath{T_\text{CO}}}

\newcommand{\muh}{\ensuremath{\mu_{\hy}}}
\newcommand{\muf}{\ensuremath{\mu_f}}
\newcommand{\nf}{\ensuremath{n_f}}

\newcommand{\hhco}{\ch{H2CO}}
\newcommand{\hy}{\ensuremath{\ch{H}}}
\newcommand{\hh}{\ensuremath{\ch{H_2}}}

\newcommand{\Rin}{\ensuremath{r_\text{in}}}
\newcommand{\Rout}{\ensuremath{r_\text{out}}}
\newcommand{\xin}{\ensuremath{x_\text{in}}}
\newcommand{\xout}{\ensuremath{x_\text{out}}}
\newcommand{\mdisk}{\ensuremath{M_\text{disk}}}
\newcommand{\exponent}{\ensuremath{\gamma}}
\newcommand{\qs}{\ensuremath{{\exponent_\text{S}}}}
\newcommand{\qh}{\ensuremath{{\exponent_\text{H}}}}
\newcommand{\qt}{\ensuremath{{\exponent_\text{T}}}}
\newcommand{\qta}{\ensuremath{{\exponent_\text{T,a}}}}
\newcommand{\qtm}{\ensuremath{{\exponent_\text{T,m}}}}
\newcommand{\qn}{\ensuremath{{\exponent_\text{n}}}}
\newcommand{\SC}{\ensuremath{\Sigma_c}}
\newcommand{\RC}{\ensuremath{r_c}}
\newcommand{\HC}{\ensuremath{H_c}}
\newcommand{\nc}{\ensuremath{n_c}}
\newcommand{\rhoc}{\ensuremath{\rho_c}}
\newcommand{\lstar}{\ensuremath{L_\star}}
\newcommand{\tmid}{\ensuremath{T_\text{m}}}
\newcommand{\tmidrc}{\ensuremath{T_\text{m,C}}}
\newcommand{\tatm}{\ensuremath{T_\text{a}}}
\newcommand{\tatmrc}{\ensuremath{T_\text{a,C}}}
\newcommand{\exsig}{\ensuremath{\gamma_S}}
\newcommand{\exdens}{\ensuremath{\gamma_n}}
\newcommand{\exh}{\ensuremath{{\gamma_H}}}
\newcommand{\extemp}{\ensuremath{\gamma_T}}
\newcommand{\extempm}{\ensuremath{\gamma_{T_m}}}
\newcommand{\extempa}{\ensuremath{\gamma_{T_a}}}

\newcommand{\Celem}{\ensuremath{[\text{C}]_{\text{elem}}}}
\newcommand{\Oelem}{\ensuremath{[\text{O}]_{\text{elem}}}}
\newcommand{\Nelem}{\ensuremath{[\text{N}]_{\text{elem}}}}
\newcommand{\Selem}{\ensuremath{[\text{S}]_{\text{elem}}}}
\newcommand{\CO}{\ensuremath{{\rm [C/O]_{elem}}}}
\newcommand{\CN}{\ensuremath{{\rm [C/N]_{elem}}}}

\def \mnras{MNRAS}
\def \pra{Phys Rev A}
\def \jcp{JCP}
\def \apj{ApJ}
\def \aap{A\&A}
\def \apjl{ApJL}
\def \nat{Nature}

\newpage
\begin{abstract}
  The spatial distribution of the chemical reservoirs in protoplanetary disks is key to elucidate the composition of planets, especially habitable ones. However, the partitioning of the main elements among the refractory and volatile phases is still elusive. Key parameters such as the carbon-to-oxygen (C/O) elemental ratio and the ionization fraction remain poorly constrained, with the latter potentially orders of magnitude lower than in the interstellar medium. Moreover, the thermal structure of the gas is also poorly known, despite its deep influence on gas-phase chemistry. In this context, ortho-to-para ratios could provide selective and sensitive probes. Recent ALMA observations have measured the spatially resolved column density of ortho- and para-H$_2$CO in the transition disk orbiting TW~Hya and derived the radial profile of the ortho-to-para ratio. Yet, current disk models do not include the nuclear-spin-resolved chemistry required to interpret these observations. The present work aims to fill this gap, by combining a parametric disk physical model of TW~Hya with the UGAN network, updated to include a comprehensive description of the nuclear-spin-resolved chemistry of formaldehyde. This new model successfully reproduces the observed column density of H$_2$CO to within a factor of 2, as well as the measured ortho-to-para ratio which varies from $\sim $1.5 in the outer disk to $\sim$3 inside 90~au. In particular the low value of this ratio beyond $\sim$90 au is well explained by our model. However, the statistical value of 3 measured below $\sim$70 au cannot be reproduced, suggesting that additional processes involving ices may be involved. Our parameter space exploration shows that the abundance of H$_2$CO is highly sensitive to the C/O elemental ratio and to the cosmic-ray ionization rate. Future observations of ortho- and para-H$_2$CO, based on well selected rotational transitions, in a large sample of disks, appear highly desirable.
\end{abstract}

\newpage
\section{Introduction}

Protoplanetary disks are the birthplaces of planetary systems, where complex physical and chemical processes govern the transformation of gas and dust into planets. A key challenge in disk studies is to establish how elements are partitioned between refractory material and volatile reservoirs (gas and ices), since this distribution sets the initial composition of forming planets, including potentially habitable ones.
Despite major improvements in the observational (ALMA, JWST, NOEMA) and modeling capabilities, the spatial composition of protoplanetary disks remains only partially understood. In fact, an array of physical and chemical processes, acting on different time scales, makes a comprehensive self-consistent modeling of such environments a challenging task. \cite{Henning2013, oberg2023}

Among the most critical open questions are the two-dimensional temperature distribution, the spatial variations of the carbon-to-oxygen (C/O) budget, and the ionization fraction. Despite having a strong influence on molecular abundances, they are difficult to constrain observationally. To overcome these limitations, one approach is to focus on specific chemical tracers, such as isotopic ratios \rlg{and nuclear-spin isomer ratios} (so-called ortho-to-para ratios, OPRs).

Historically, OPRs have been used to infer the formation temperature of molecules in astrophysical environments. \citep{mumma_chemical_2011} However, it is now clear that they rather reflect the current physical and chemical conditions in the gas. \cite{hama_ortho--para_2018, Faure2019} Still, OPRs are powerful tools because they are governed by spin-resolved branching ratios that obey rigorous nuclear-spin selection rules. \cite{Rist2013} Therefore, OPRs provide highly selective and accurate probes of physical and chemical conditions, especially temperature, ionization fraction, and chemical routes. \cite{Faure2013, Legal2014, HB2018, Faure2019}

Formaldehyde (\ce{H2CO}) is a particularly valuable molecule in this context. It is both a common tracer in disks and a key intermediate in the chemistry leading to complex organic molecules. The nearby transition disk TW~Hya, at a distance of only 59 pc, provides an ideal system in which to study the spin-chemistry of \ce{H2CO}. As one of the nearest and most extensively observed protoplanetary disks, TW~Hya benefits from high-resolution observations that tightly constrain its physical and chemical structures. Its well-characterized temperature and density profiles, combined with a relatively low disk mass, make it an ideal testbed to benchmark disk chemistry models. While the first measurement of an \ce{H2CO} OPR in a protoplanetary disk was reported by \citet{guzman2018} toward HD~163296 using SMA and ALMA, more recent ALMA observations have spatially resolved the \hhco\ OPR in the TW~Hya disk. \cite{Terwisscha_van_Scheltinga_2021} These measurements open new means of investigating the physical and chemical conditions prevailing in protoplanetary disks, but interpreting them requires models that explicitly account for spin-resolved chemistry, a capability that existing disk models lack.

In this work, we address this gap by developing the first protoplanetary disk model that incorporates nuclear-spin chemistry for \ce{H2CO}. Using a parametric physical model of the TW~Hya disk (presented in Section \ref{sec:methods}) in combination with the UGAN chemical network, extended to include a detailed description of formaldehyde spin chemistry (as described in Section \ref{sec:results}), we investigate the radial abundance and column density distribution of ortho- and para-\ce{H2CO} and compare our predictions with ALMA measurements (see Section \ref{sec:ccl}). Perspectives in the field of spin chemistry applied to disks are given in the last Section.

\section{Methods}
\label{sec:methods}

The chief objective of the present work is to compute the spatial distribution of the ortho- and para-\ce{H2CO} molecules in proto-planetary disks. To this aim, we have developed a steady-state parametric physical model for the density and temperature of the gas. Combined with the \texttt{UGAN} chemical network, \citep{HB2018} this allows us to compute the abundance of both symmetries of \ce{H2CO} and compare them with observations in the TW~Hya disk.

\subsection{The Physical Disk Model}
\label{sec:phys}

Our modeled disk has the commonly adopted azimuthal symmetry with a purely ortho-radial Keplerian velocity field, $v(r)=(G\mstar/r)^{1/2}$, with \mstar\ the mass of the central protostar, and $r$ the radius. \cite{andrews2012, rosenfeld2013, cleeves2015, legal2019, lee2021} The disk is radially truncated between \Rin\ and \Rout. The radial profile of the mass density is dictated by the mass of the disk, \mdisk, and by the radial profile of the surface density which assumes a power-law dependence, $\Sigma(r) = \SC (r/\RC)^{-\qs}$, with \SC\ the value at the characteristic radius \RC, which is fixed by the normalization condition on the disk mass. The vertical mass density profile derives from balancing the pressure gradient and gravity, while adopting a perfect gas equation of state. Assuming that the vertical temperature gradient gives negligible contribution to the pressure gradient, the vertical density profile reduces to the hydrostatic one and a scale height is obtained, which depends only on the midplane temperature, \tmidrc: $H(r) = \HC (r/\RC)^\qh$ with $\HC \propto (\tmidrc/\mstar)^{0.5}$. This assumption is justified by the fact that most of the mass is located close to the midplane. Finally, the mass density structure is given by $\rho(r,z) = (\Sigma(r)/H(r)/\sqrt{2\pi}) \exp(-z^2/2H^2(r))$ and is related to the number density of H nuclei $\nh = n(\hy) + 2n(\hh)$ through $\rho(r,z) = \nh(r,z) \muh \amu$ with \amu\ the atomic mass unit and $\muh = 1 + 4y$ where $y=n(\ch{He})/\nh = 0.1$ is the present-day abundance of helium in the solar neighborhood. The number density of H nuclei is thus given by $\nh(r,z) = \nc (r/\RC)^{-\qn} \exp[-z^2/2H^2(r)]$, where $\nc=\SC/(\sqrt{2\pi} \HC \mu \amu)$ and $\qn = \qh + \qs$. We have also considered another common analytical description of the surface density which includes an exponential taper at large radii: \citep{andrews2012, lee2024} $\Sigma(r) = \SC (r/\RC)^{-\qs} \exp[-(r/\RC)^{2-\qs}]$.

Although the vertical temperature variations are neglected in deriving the density profile, they are not for the chemical calculations. The adopted vertical temperature profile models a continuous transition from a midplane temperature $\tmid(r)$ to the atmospheric value in the upper layers of the disk, $\tatm(r)$. In the present work, we adopt the following formulation,
\begin{equation}
  \label{eq:t}
  T(r,z) = \tmid(r) + \frac{\tatm(r)-\tmid(r)}{2} \lc 1 + \tanh \pfrac{z-z_0}{\Delta(r)} \rc.
\end{equation}
This is similar to a commonly used parametric profile, \citep{dartois2003, andrews2012, legal2019} while consisting in a single function, with two parameters, $z_0$ and $\Delta$, providing direct and independent control of the location and width of the transition from the midplane to atmospheric values. The hydrostatic assumption rests upon $z_0/H$ being substantially larger than unity, and most models in the literature take $z_0/H \approx 3-4$. As is customary in parametric disk models, the midplane and atmospheric temperature profiles are radial power-laws, $\tmid(r) = \tmidrc(r/\RC)^{-\qtm}$ and $\tatm(r) = \tatmrc (r/\RC)^{-\qta}$. The exponent \qtm\ is related to the scale height exponent through $\qh = (3-\qtm)/2$. The vertical temperature profile may, however, be much more complex than a single transition from a low to a high temperature. In fact, more physically based temperature distributions \citep{lee2024} suggest that the gas temperature is better modeled with a two-layers profile. Therefore, we have considered a double-tanh temperature profile consisting of the sum of two profiles of the form given by Eq.\,\eqref{eq:t} (see Model 3 in Table\,\ref{tab:param}):
\begin{equation}
    \label{eq:double}
    T(r,z) = \tmid(r) + \frac{\tatm(r)-\tmid(r)}{2} \lc 1 + \tanh \pfrac{z-z_0}{\Delta(r)} \rc + \frac{\tatm'}{2} \lc 1 + \tanh \pfrac{z-z_0'}{\Delta'(r)} \rc,
\end{equation}
where primed quantities refer to the upper layer. In what follows, we thus consider three models summarized in Table\,\ref{tab:models}, and the free parameters are listed in Table \ref{tab:param}. The parameters were obtained manually so as to reproduce the temperature profile of \citet{andrews2012} for Models 1 and 2, and of \citet{lee2024} for Model 3.

\begin{table}[t]
  \centering
  \caption{Three Types of Models Considered in This Work (see also Table\,\ref{tab:param}). For the \af{gas-phase} elemental abundances \af{of carbon and oxygen} (always given relative to H nuclei), a distinction is made above and below the CO snow surface. \af{The gas-phase elemental abundance of nitrogen is fixed at half the ISM value, i.e. 3.2\tdix{-5},} and the \af{gas-phase elemental} sulfur abundance is 8.0\tdix{-8} \cite{Legal2014}. See Section~\ref{sec:models} for details.}
  \label{tab:models}
  \begin{tabular}{ll ccc}
    \toprule
    && Model 1 & Model 2 & Model 3 \\
    \midrule
    Temperature$^a$ && Single & Single & Double \\
    Tapered density$^b$ && No & Yes & Yes \\
    \depc$^c$ & above & 30 & 30 & 30\\
    & below & 300 & 300 & 3000$^d$\\
    \depn$^e$ &  & 1.5 & 1.5 & 1.5\\
    \CO$^f$
    && 1 & 1 & 1 \\
    && 1 & 1 & 1 \\
    \CN$^g$ 
    && 6.5\tdix{-2}& 6.5\tdix{-2}& 6.5\tdix{-2}\\
    && 6.5\tdix{-3}& 6.5\tdix{-3}& 6.5\tdix{-4}\\
    \bottomrule
  \end{tabular}
  \tabnotes $^a$ Temperature profile using a single or double tanh as in Eq.\,\eqref{eq:t}. $^b$ Whether a tapered density profile is used or not. $^c$ Depletion factor $\depc=8.3\tdix{-5}/\Celem$ above (first row) and below (second row) the snow surface. $^d$ A higher depletion is applied to reproduce the CO column density profile of \citet{huang2018}. \af{$^e$ Depletion factor $\depn=6.4\tdix{-5}/\Nelem$.}
  $^f$ The gas-phase elemental ratio above and below the snow surface. The elemental abundance of oxygen is computed as $\Oelem = \Celem/\CO$. $^g$ The elemental \CN\ ratio above and below the snow surface.
\end{table}

\begin{table}[t]
  \centering
  \caption{Physical Parameters Used for Our Three Disk Models of TW Hya}
  \label{tab:param}
  \begin{tabular}{lccccc}
    \toprule
    Parameters & unit & Fiducial & Model 1 & Model 2 & Model 3$^a$ \\
    \midrule
    \mstar   & \msol & 0.8$^b$ \\
    \mdisk   & \msol & 0.023$^c$ \\
    $y$      &       & 0.1$^d$\\
    \Rin     & \au   & 4\\
    \Rout    & \au   & 200\\
    \RC      & \au   & 35\\
    \HC      & \au   &    & 3.5   & 3.5   & 3.5 \\
    \qh      &       &    & 1.325 & 1.325 & 1.3$^e$  \\
    \SC      & g\cc &    & 5.2   & 11.0  & 11.0 \\
    $\qs$    &       &    & 2.5   & 1.0   & 1.0  \\
    \nc&$10^{10}$\ccc&    & 1.7   & 3.6   & 3.6  \\
    \tmid    & K     &    & $16 (r/\RC)^{-0.35}$ &  $16 (r/\RC)^{-0.35}$ &  $16 (r/\RC)^{-0.35}$ \\
               &&&&& 0 \\
    \tatm    & K     &    & $54 (r/\RC)^{-0.5}$ & $54 (r/\RC)^{-0.5}$ & $25 (r/\RC)^{-0.3}$\\
    $\tatm'$ &&&&&  $400$\\
    $z_0$    & \au   &   & 1.5 & 1.5 & 0.4 \\
    $z_0'$   & \au   &&&& 3.5 \\
    $\Delta(r)$ & \au  &   & 1.0 & 1.0 & 0.2 \\
    $\Delta'(r)$ & \au &&&& $(r/\RC)^{0.15}$\\
    \bottomrule
  \end{tabular}
  \tabnotes $^a$ For the temperature in Model 3, the expressions for \tmid, \tatm, $z_0$, and $\Delta(r)$ refer to the lower and primed quantities are for the upper layer.  $^b$ From \citet{canta2021, qi2013}. $^c$ From \citet{kama2016}. $^d$ $y=n(\ch{He})/\nh$. $^e$ Exponent taken from \citet{lee2024}.
\end{table}

The hydrostatic vertical density profile allows column densities to be computed with high accuracy using the error function, $\erfc(z) = 2/\sqrt{\pi} \int_z^\infty e^{-t^2} dt = 1-\erf(z)$. We thus readily obtain the H nuclei column density above a given height $z$ as:
\begin{equation}
  \label{eq:NH}
  \NH(r,z) = \sqrt{\frac{\pi}{2}} \nc \HC \pfrac{r}{\RC}^{-\qs} \erfc\pfrac{z}{\sqrt{2} H(r)}.
\end{equation}
In the above calculation, only one side of the disk is considered.
The column density may then be converted in a visual extinction through \cite{wagenblast1989} $\NH [\cc] = A_V \times 1.6\times10^{21}$.

\begin{figure}
  \centering
  \def\ww{.49\hsize}
  \includegraphics[width=\ww]{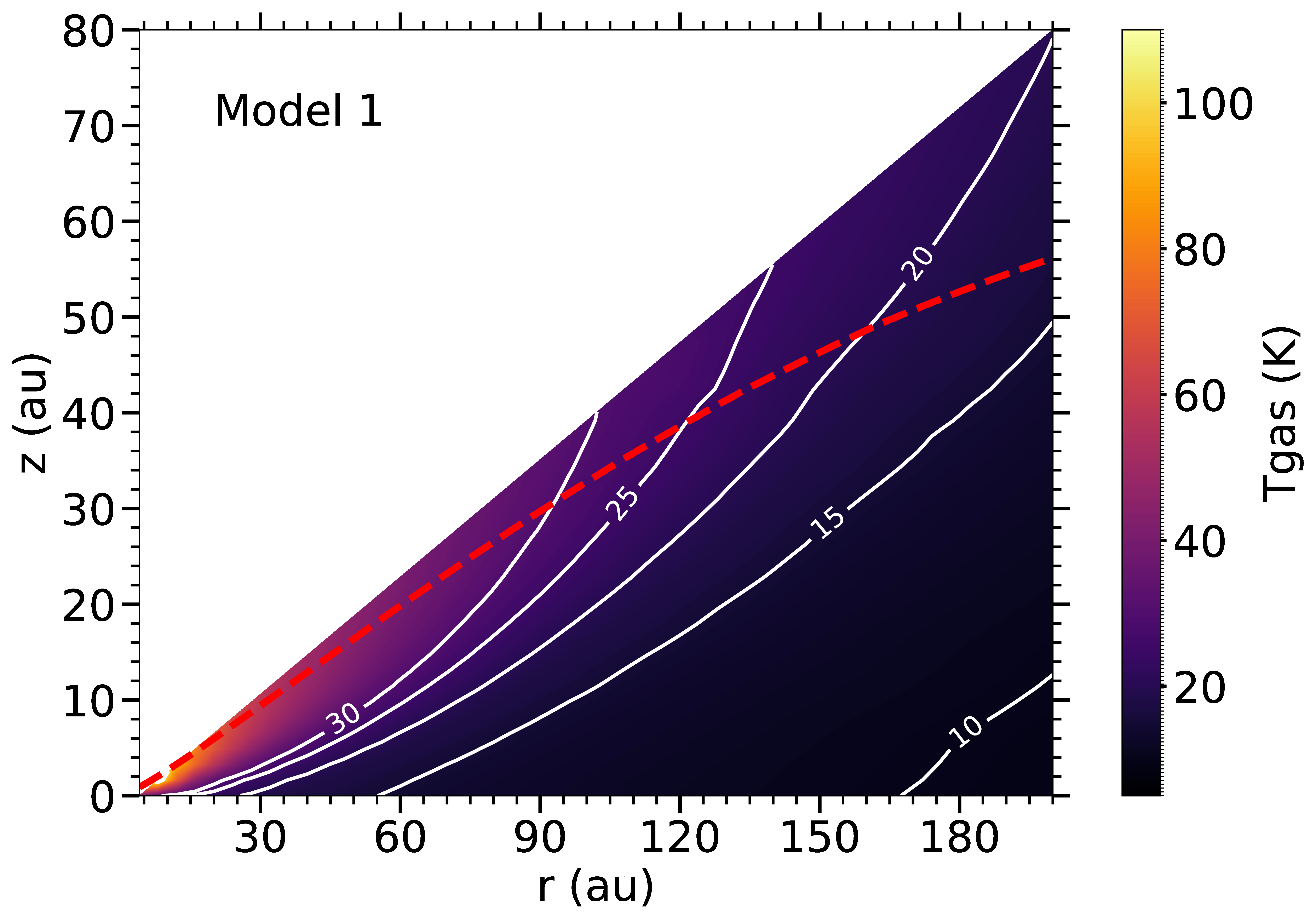}\hfill%
  \includegraphics[width=\ww]{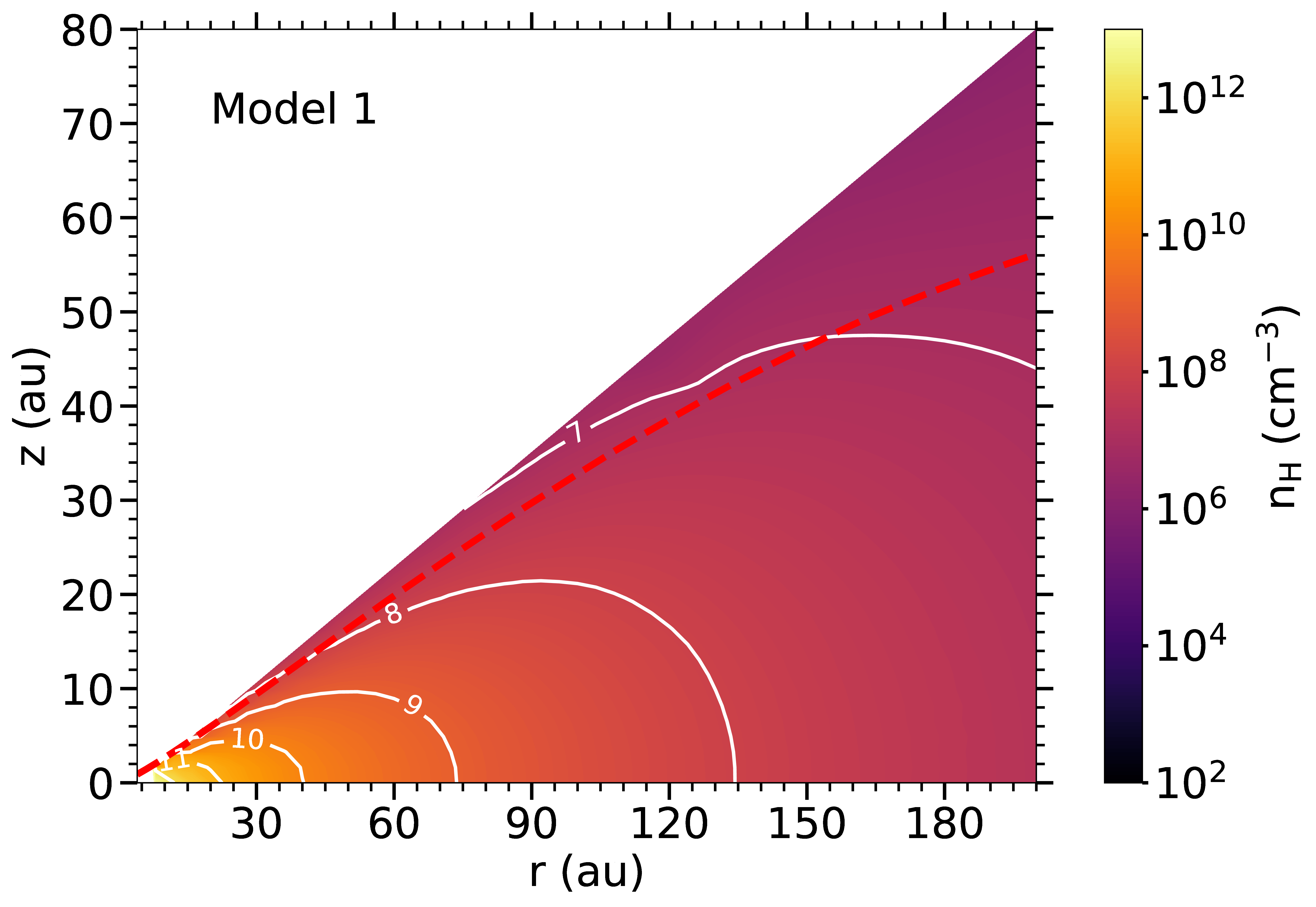}\smallskip\\
  \includegraphics[width=\ww]{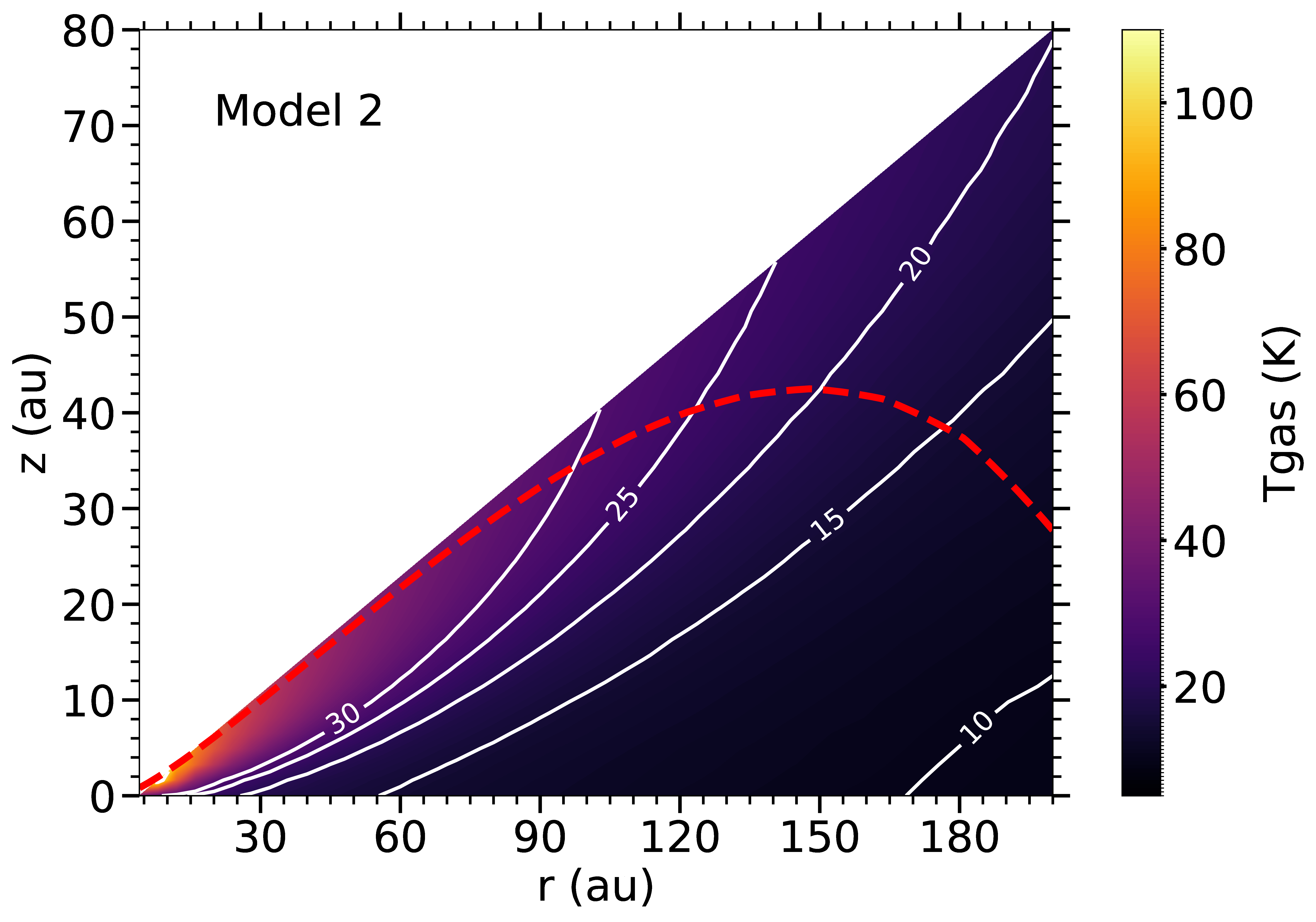}\hfill%
  \includegraphics[width=\ww]{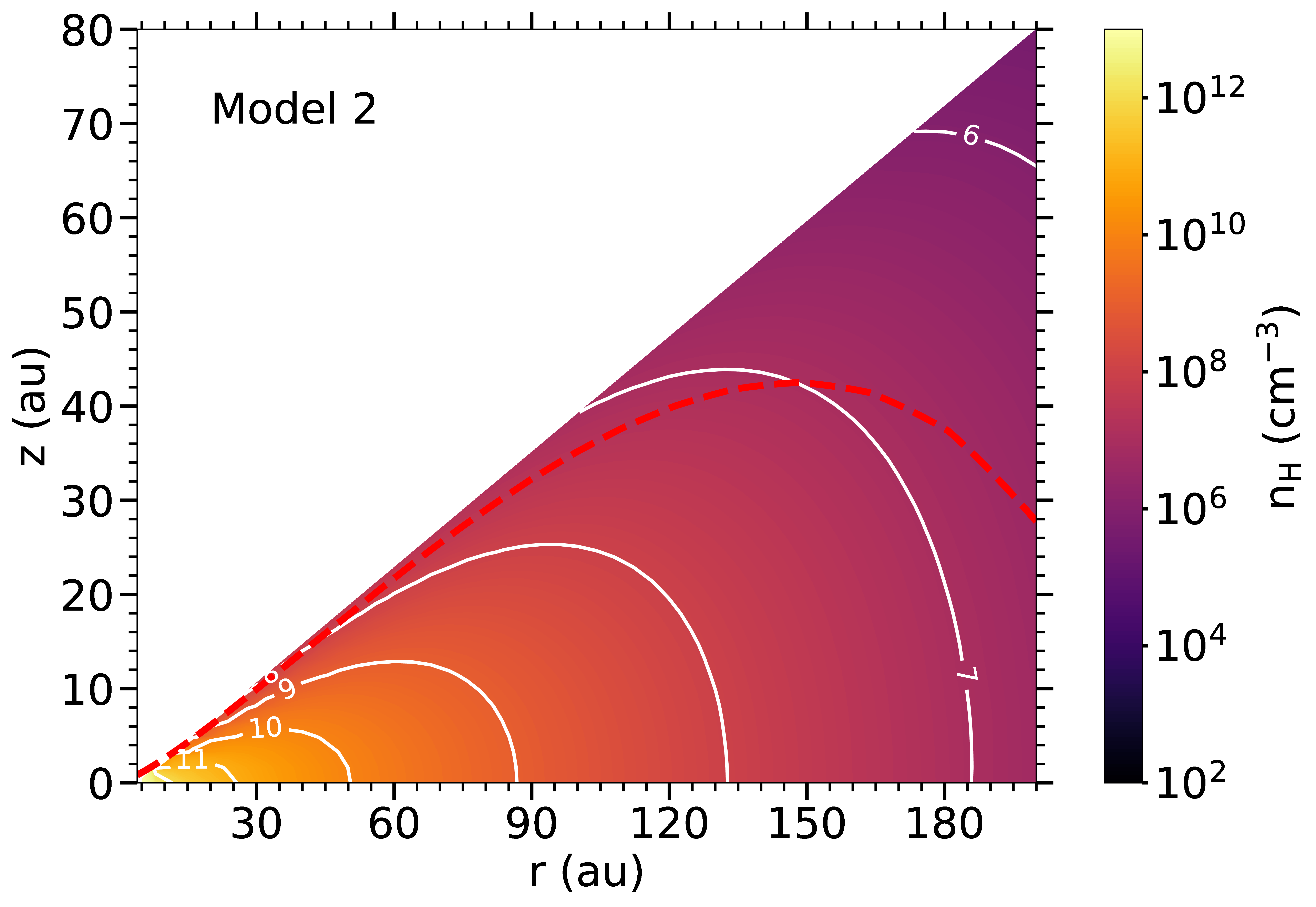}\smallskip\\
  \includegraphics[width=\ww]{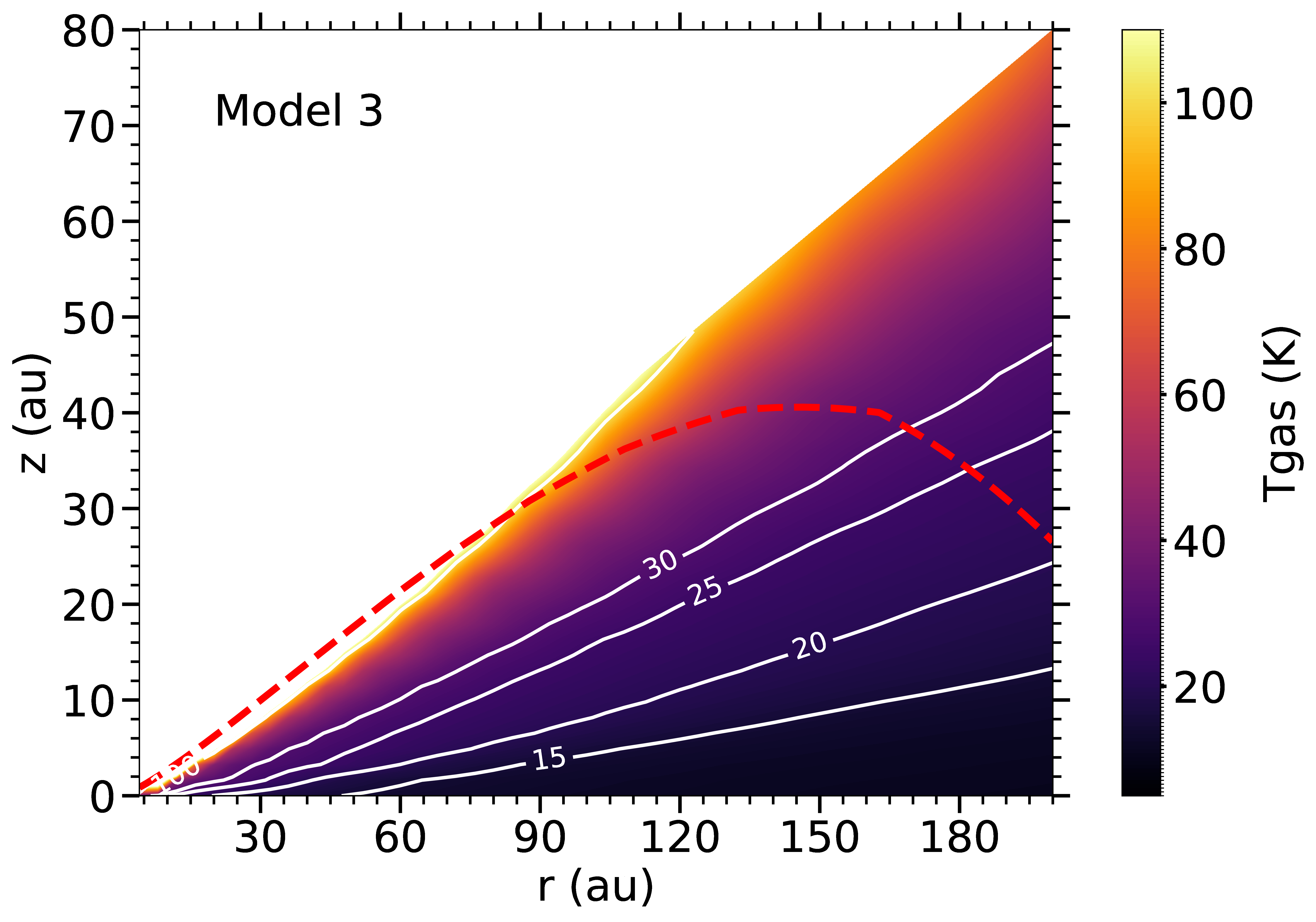}\hfill%
  \includegraphics[width=\ww]{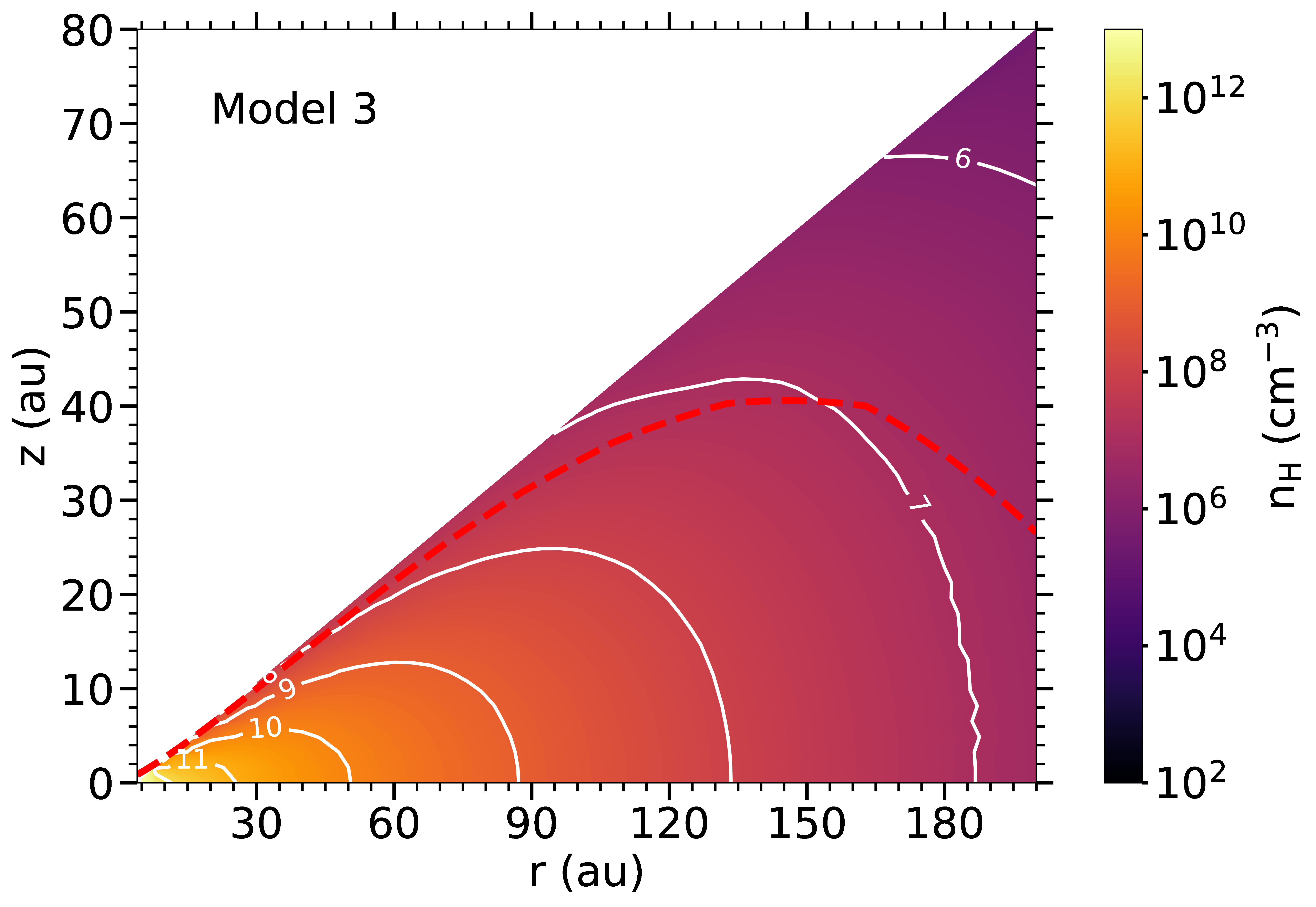}
  \caption{Disk physical structure from our three parametric models (Models 1 to 3 from top to bottom, see Tables \ref{tab:models} and \ref{tab:param}). The 2D gas temperature (left) and total H density (right) are shown as colors, with iso-curves in white. The dashed red line corresponds to $\av=1$~mag (as measured from the top).}
  \label{fig:phys_map}
\end{figure}

\phb{The three physical models considered in this work have been constructed as follows. For all models, the physical structure of \citet{lee2024} (based on \citet{kama2016})} was taken as a reference model. The power-law density profile in Model 1 is constrained by the disk mass between 4 and 200 au, and reproduces the reference density to within a factor of 2 from 10 to 165 au, and overestimates it by a factor of 4 at 200~au \af{(see Fig.~S1 in Supporting Information)}. 
Our midplane temperature power-law reproduces the reference values to better than 2~K between 10 and 150 au, and is 15~K lower at 200 au. The midplane temperature profile is common to all three models. The atmospheric temperature profile of Models 1 and 2 closely follows the parametric profile of \citet{andrews2012} In contrast, the two-layers temperature profile of Model 3 was fitted manually against the model of \citet{lee2024}, providing very good agreement up to $z/H\sim 3$ within 90 au, and up to $z/H\approx 2.5$ at larger radii \af{(see Fig.~S1 in Supporting Information)}. The 2D physical structures corresponding to models 1 to 3 are shown in Fig.\,\ref{fig:phys_map}.

\subsection{Ortho-Para Chemistry of Formaldehyde in UGAN}

\af{The \texttt{UGAN} nuclear-spin-resolved chemical network was originally developed from the work of \citet{Flower2006a} (hereafter F06) devoted to the chemistry of prestellar cores. It includes species containing H, D, He, C, N, O, S and separates the nuclear-spin states of \ch{H_2}, \ch{H_2^+}, \ch{H_3^+} (including deuterated isotopologues) and those of carbon, nitrogen, oxygen and sulfur hydrides. A first update of F06 consisted in a revision of the nitrogen hydrides chemistry was presented in \citet{Legal2014} The nuclear-spin selection rules were derived using the method of Oka, \cite{oka2004} which is based on the conservation of the rotational symmetry of the nuclear-spin isomers. The second update by \citet{HB2018} (hereafter HB18) consisted of extending the work of \citet{Legal2014} to the entire F06 network in a systematic manner for all hydrides containing C, N, O, and S atoms, along with their deuterated forms for N- and O-hydrides. The nuclear-spin selection rules were derived from the permutation symmetry approach of Quack \cite{quack1977}, which is more general than the method of Oka \cite{oka2004} and adapted to deuterium nuclei. The selection rules depend on the reaction mechanism, with two limiting cases: hopping (direct exchange) and full scrambling (indirect exchange) of protons. In HB18 and in this work}, full scrambling is assumed because at low temperatures (i.e. below $\sim$ 100~K), reactions generally proceed via long-lived intermediate complexes where complete randomization of identical nuclei takes place. Direct evidence of nuclear-spin selection rules in a reaction involving a long-lived complex was in fact observed only recently using ultracold experimental techniques. \cite{Hu2021} On the other hand, we note that ion-trap measurements on the deuterated isotopic variants of the \ce{HCl^+ + H2 -> H2Cl^+ + H} reaction indicate that this reaction proceeds via proton-hop, with no evidence of scrambling, \citep{Jimenez2025} in agreement with classical trajectory calculations. \cite{legal2017} It should be noted that the proton-hop scenario would result in statistical nuclear-spin ratios, i.e. an OPR of 3 for \ce{H2CO}, as predicted for the deuterated isotopologues of ammonia. \citep{Harju2025} 
The third update by \citet{Faure2019} consisted of a revision of the oxygen hydrides chemistry, and particularly that of water. Subsequent targeted updates to the \texttt{UGAN} network were published in \citet{HB2020,HB2022} Many reaction rate coefficients were also updated from a literature survey.

We describe below the fourth major update of UGAN, which primarily involves a revision of the carbon hydrides chemistry and the introduction of the ortho-para chemistry of formaldehyde. Specifically, we have:
\begin{itemize}
    \item added \ce{H2CO}, \ch{HCO}, \ch{H_2CO^+} and \ch{H_2COH^+} as new gas-phase reactive species.
    \item updated the nuclear-spin-resolved chemistry of \ce{CH3} and O($^3P$), the key precursors of \ce{H2CO}.
    \item implemented revised branching ratio for \ce{CH5+} dissociative recombination based on CRYRING storage-ring experiments \cite{Kaminska2010}.
    \item incorporated low-temperature neutral-neutral destruction pathways of \ce{H2CO} with CN and \ce{C2H}, constrained by recent CRESU (Cinétique en Écoulement Supersonique Uniforme) measurements \cite{West2023,Douglas2024}.
    \item adopted new dissociative recombination rates for \ce{H2COH+} \cite{Hamberg2007}, with branching ratios consistent with observational constraints \cite{Bacmann2016}.
\end{itemize}

\subsubsection{Formation Pathways}

\af{The chemistry of \ce{H2CO} can follow three distinct routes: (i) ion-neutral gas-phase chemistry, (ii) neutral-neutral gas-phase chemistry and (iii) surface chemistry. The first route should proceed via the radiative association of \ch{HCO^+} with \ch{H_2} or H, followed by dissociative recombination or charge exchange, but this route is unlikely to be dominant. \citep{Klemperer2010} The primary neutral-neutral \phb{formation} pathway in the gas-phase should be the reaction between the radical \ch{CH_3} and atomic oxygen O($^3P$),
\begin{equation}
  \ce{CH3 + O -> H2CO + H} \label{eq:h2co}
\end{equation}
\phb{Finally, efficient formation of \ce{H2CO} by the hydrogenation of CO in \ce{H2O}$-$CO ice was demonstrated experimentally} at 10~K using an atomic hydrogen beam experiment. \cite{Watanabe2002} The second and third routes should therefore compete in cold (hereafter ‘cold’ refers to kinetic temperatures below 100~K)} environments, but the variations of their relative importance with varying physical conditions has yet to be established. In particular, the ice route requires an efficient thermal or nonthermal desorption mechanism for releasing intact \ce{H2CO} into the gas-phase. Despite experimental efforts, such a mechanism has so far remained elusive. \cite{Faure2025} 

\af{In this work, the second and third routes are considered, with
  surface chemistry being treated only implicitly by assuming that the
  grains have a static ice mantle composed of \ce{H2O}, CO, \ce{CO2},
  \ce{H2CO}, etc. (their abundances relative to \ce{H2O} can be derived from} Table~3 of HB18) with a canonical \ce{H2CO}
  abundance of 6\% with respect to water ice. \citep{Boogert2015} Thus, there are ices everywhere in the molecular layer but adsorption and depletion are not treated explicitly: the abundances of solid species (relative to H nuclei) are set by the depletion factors that vary above and below the snow surface (see Section~\ref{sec:models}), while their relative abundance with respect to solid water (from HB18) remain fixed. As a result, the ice and gas phases are treated as separate reservoirs, and the OPR and abundance of gaseous formaldehyde results exclusively from gas-phase reactions. The major updates of \texttt{UGAN} thus concern the \phb{nuclear-spin-resolved} chemistry (formation and destruction) of \ch{CH_3}, O$(^3P)$, and \ce{H2CO}. We also note that the OPR of \ce{H2CO} in ice should equal the statistical value of 3, as supported by theory and laboratory experiments (see \citet[and references therein]{yocum2023}).

\af{Carbon hydrides chemistry in cold environments starts with the slow radiative association:
\begin{equation}
  \ce{C+ + H2 -> CH2+ + h\nu,}
  \label{eq:ch2p}
\end{equation}
followed by an exothermic and fast hydrogen abstraction leading to the pivotal \ce{CH3+} ion: 
\begin{equation}
  \ce{CH2+ + H2 -> CH3+ + H},
  \label{eq:ch3p}
\end{equation}
and, finally, another slow radiative association with \ce{H2}:
\begin{equation}
  \ce{CH3+ + H2 -> CH5+ + h\nu},
  \label{eq:ch5p}
\end{equation}
forming the radical \ce{CH3}, among other products, after dissociative recombination:
\begin{align}
    \ce{CH5+ + e-} & \ce{-> CH4 + H}, \label{eq:ch3_1}\\
    \ce{} & \ce{-> CH3 + H2},\\
    \ce{} &\ce{-> CH3 + 2H},\\
    \ce{} &\ce{-> CH2 + H2 + H},\\
    \ce{} &\ce{-> CH2 + 2H2} \label{eq:ch3_2}.
\end{align}
The above set of reactions should dominate the formation of \ch{CH_3} in cold environments. An important update of \texttt{UGAN} concerns reactions~(\ref{eq:ch3_1})-(\ref{eq:ch3_2}) where the thermal rate coefficient and branching ratios were taken from the CRYRING storage-ring measurements of \citet{Kaminska2010} We note that a key reaction regarding the ortho-para chemistry of \ch{CH_3} (and thus \ce{H2CO}) is the nuclear-spin thermalization\footnote{\af{Thermalization here refers to \ch{CH_3^+} reaching a nuclear-spin equilibrium via proton scrambling, which is not necessarily full thermal equilibrium.}} of \ch{CH_3^+} by \ch{H_2}, which will set the OPR of \ch{CH_5^+} and then \ch{CH_3} (see Supporting Information). Such thermalization processes are typically} neglected in the \texttt{UGAN} network, except the fast equilibration reactions of \ch{H_2} with H$^+$, \ch{H_3^+} and \ch{HCO^+} (and their deuterated variants). However, efficient proton scrambling should occur in the protonated methane (\ch{CH_5^+})$^*$ complex, as supported by the fast isotopic H-D exchanges observed in the reaction \ce{CH3+ + HD -> CH2D+ + H2} (see \citet{Gerlich2005} and references therein). Thermalization by \ch{H_2} should thus drive the OPR of \ch{CH_3^+} toward thermal equilibrium above a critical temperature of $\sim 15-20$~K \citep{Faure2013}, where OPR(\ch{CH_3^+}) should converge toward the statistical value of 1. The thermalization reaction of \ch{CH_3^+} by \ch{H_2} was added in our network following the state-to-state statistical approach of \citet{Rist2013}

\af{Oxygen atoms react primarily with H$^+$ to create O$^+$ ions (by charge transfer) or with \ch{H_3^+} to form \ch{OH^+} and \ch{H_2O^+}:
\begin{equation}
\begin{aligned}
    \ce{O + H3+} & \ce{-> OH+ + H_2} \\
    \ce{} & \ce{-> H2O+ + H}.
\end{aligned}
    \label{eq:h2op}
\end{equation}
The above reaction was revisited recently using merged-beam measurements by \citet{Hillenbrand2022} Thermal rate coefficients for the two product channels were derived by these authors for the temperature range from 10 to 1000~K, and a temperature-dependent branching ratio was observed. We adopted the thermal rate coefficients where the fine-structure level populations of O($^3P_J$) are assumed in thermal equilibrium (see Table~2 of \citet{Hillenbrand2022}).} 

\phb{Finally, as explained above, formaldehyde should predominantly form via the reaction of \ce{CH3} with atomic oxygen (reaction \eqref{eq:h2co}), for which a competing channel is
\begin{equation}
  \ce{CH3 + O -> CO + H2 + H}.\label{eq:h2co_2}
\end{equation}}
\phb{The most recent kinetic measurements for the \ce{CH3 + O} reaction report a total rate coefficient of $\sim 1-2 \times 10^{-10} \cccs$ over the temperature range $259-2580$~K, being $1.26 \times 10^{-10}$\cccs\ at room temperature, and} \af{with a temperature-independent branching ratio of 55\% for \ce{H2CO + H} and 45\% for \ce{CO + H2 + H} \cite{Hack2005}, in very good agreement with the calculations of \citet{Xu2015} We adopted the rate coefficient and branching ratio of \citet{Hack2005} with no temperature dependence. We note, however, that measurements below 100~K would be highly beneficial.}

\subsubsection{Destruction Pathways}

\phb{Destruction of formaldehyde is dominated by reactions with the abundant CN and \ce{C2H} radicals and with \ce{HCO+}.  The rate coefficients for the destruction of \ce{H2CO} by \ce{C2H} and CN} \af{were measured recently down to 22~K using the CRESU technique: \citep{West2023,Douglas2024}
\begin{align}
  \ce{H2CO + CN} & \ce{-> HCN + HCO}, \label{eq:cn}\\
  \ce{H2CO + C2H} & \ce{-> C2H2 + HCO}, \label{eq:c2h_1}\\
                 & \ce{-> C2H2 + CO + H}. \label{eq:c2h_2}
\end{align}

These reactions belong to a class of neutral-neutral processes that accelerate below $\sim$200~K, where the temperature dependence changes from positive to negative. The postulated mechanism involves the formation of a weakly bound van der Waals complex, whose lifetime is prolonged at low temperature, and tunneling through a small activation barrier. \citep{Heard2018} This is indeed supported by {\it ab initio} calculations which can be fitted to the experimental data to provide rate coefficients over a large temperature range. The recommended best-fit expressions from \citet{West2023} and \citet{Douglas2024} for reactions with CN and \ce{C2H}, respectively, were adopted in \texttt{UGAN}. Another key destruction pathway of \ce{H2CO} is the protonation reaction with \ch{HCO^+}:
\begin{equation}
  \ce{H2CO + HCO+ -> H2COH+ + CO}.
  \label{eq:hcop}
\end{equation}
For this reaction, and other similar ion-molecule reactions involving \ce{H2CO} or HCO, we used the capture theory of Su \& Chesnavich \cite{Su1982} for ion-polar molecule collisions. We note, in particular, that the formaldehyde cation \ce{H2CO+} can form via the charge transfer reaction \ce{H2CO + H+ -> H2CO+ + H} and via the protonation of HCO by e.g. \ce{HCO+}. While there is no measurement for the dissociative recombination (DR)} of \ce{H2CO+} (we use the KIDA \citep{Wakelam2024} recommendation), the DR of \ce{H2COH+} was measured at the CRYRING storage-ring by: \citet{Hamberg2007}     
\begin{equation}
  \begin{aligned}
    \ce{H2COH+ + e-} & \ce{-> CO + H + H2}, \\
    \ce{} & \ce{-> HCO + H + H},\\
    \ce{} & \ce{-> H2CO + H},\\
    \ce{} &\ce{-> CH + H2O}, \\
    \ce{} &\ce{-> CH2 + OH}.
  \end{aligned}
  \label{eq:h2cohp}
\end{equation}
The rate coefficient and partial branching ratios were taken from, \citet{Hamberg2007} with the additional assumption that the channels preserving the C-O bond (experimental branching ratio of 92\%) are distributed among CO (31\%), HCO  (30\%) and \ce{H2CO} (31\%). The branching ratio for the HCO channel (30\%) is the upper limit derived by \citet{Bacmann2016} from the HCO and \ch{H_2COH^+} column densities measured in the prestellar core L1689B. 

\af{The rate coefficients for the updated and new reactions in \texttt{UGAN} are listed in Table~\ref{tab:reactions} in the form of Arrhenius-Kooij fits.}

\begin{table}[t]
\centering
\small
\caption{Summary of Key Updated and New Reaction Rate Coefficients Adopted in the \texttt{UGAN} Network}\label{tab:reactions}
\begin{tabular}{@{}l c l c c c c l@{}}
\toprule
Reactants & & Products 
 & $\alpha$ [\cccs] & $\beta$ & $\gamma$ [K] & Reference & Comment \\
\midrule
\ch{CH_5^+ + e^-} & $\rightarrow$ & \ch{CH_4 + H} & $5.34(-08)$ & $-0.72$ & $0$  & (1) & Update \\
\ch{} & $\rightarrow$ & \ch{CH_3 + H_2} & $5.23(-08)$ & $-0.72$ & $0$  & (1) & Update \\
\ch{} & $\rightarrow$ & \ch{CH_3 + H + H} & $7.61(-07)$ & $-0.72$ & $0$  & (1) & Update \\
\ch{} & $\rightarrow$ & \ch{CH_2 + H_2 + H} & $1.88(-07)$ & $-0.72$ & $0$  & (1) & Update \\
\ch{} & $\rightarrow$ & \ch{CH + 2 H_2} & $3.60(-08)$ & $-0.72$ & $0$  & (1) & Update \\
\ch{O + H_3^+} & $\rightarrow$ & \ch{OH^+ + H_2} & $4.65(-10)$ & $-0.14$ & $0.67$  & (2) & Update \\
\ch{} & $\rightarrow$ & \ch{H_2O^+ + H} & $2.08(-10)$ & $-0.40$ & $4.86$  & (2) & Update \\
\ch{CH_3 + O} & $\rightarrow$ & \ch{H_2CO + H} & $6.94(-11)$ & $0$ & $0$  & (3) & New \\
\ch{} & $\rightarrow$ & \ch{CO + H_2 + H} & $5.68(-11)$ & $0$ & $0$  & (3) & New \\
\ch{H_2CO + CN} & $\rightarrow$ & \ch{HCN + HCO} & $3.72(-12)$ & $-1.09$ & $5.2$  & (4) & New \\
\ch{H_2CO + C_2H} & $\rightarrow$ & \ch{C_2H_2 + HCO} & $2.38(-11)$ & $-0.41$ & $-3.0$  & (5) & New \\
\ch{} & $\rightarrow$ & \ch{C_2H_2 + CO + H} & $2.51(-11)$ & $-0.37$ & $-3.54$  & (5) & New \\
\ch{H_2CO + HCO^+} & $\rightarrow$ & \ch{H_2COH^+ + CO} & $2.87(-09)$ & $-0.46$ & $0$  & This work & New \\
\ch{H_2COH^+ + e^-} & $\rightarrow$ & \ch{CO + H + H_2} & $2.39(-07)$ & $-0.78$ & $0$  & (6) & New \\
\ch{} & $\rightarrow$ & \ch{HCO + H + H} & $2.31(-07)$ & $-0.78$ & $0$  & (6) & New \\
\ch{} & $\rightarrow$ & \ch{H_2CO + H} & $2.39(-07)$ & $-0.78$ & $0$  & (6) & New \\
\ch{} & $\rightarrow$ & \ch{CH + H_2O} & $1.54(-08)$ & $-0.78$ & $0$  & (6) & New \\
\ch{} & $\rightarrow$ & \ch{CH_2 + OH} & $4.62(-08)$ & $-0.78$ & $0$  & (6) & New \\
\bottomrule
\end{tabular}
\tabnotes The rate coefficients are of the form $k(T)=\alpha \, (T/300)^\beta \exp(-\gamma/T)$. Fits are reliable over the kinetic temperature range $T=10-100$~K. Numbers in parentheses are powers of ten. References: (1) \citet{Kaminska2010}; (2) \citet{Hillenbrand2022}; (3) \citet{Hack2005}; (4) \citet{West2023}; (5) \citet{Douglas2024}; (6) \citet{Hamberg2007}
\end{table}

\subsection{Elemental Gas-Phase Abundances}
\label{sec:models}

\phb{The partitioning of the main elements (carbon, nitrogen, oxygen, sulfur) among the solid (refractory grains) and volatile (gas, ice) phases is a central, yet still elusive, question in disk chemical modeling. \cite{lee2024, oberg2023} Moreover, the relative amount of volatile carbon and oxygen, the \CO\ ratio, is known to have a dramatic impact on the chemical abundances. \citep{Legal2014,legal2019b} In the context of the present work, the \CO\ ratio is anticipated to play an important role in \ce{H2CO} chemistry. Because CO is the main reservoir of carbon, photodissociation of CO, as well as adsorption to and desorption from ices, are essential processes for disk chemistry. The photodissociation front defines the upper limit of the molecular layer, above which CO is efficiently destroyed by UV photons, as in Photon Dominated Regions (PDRs). In this work, the photodissociation front is defined as the region where the visual extinction (increasing downward, see Eq.\,\eqref{eq:NH}) is equal to 1 mag. \cite{qi2024} The molecular layer extends down to the snow surface, defined as the $T=25$~K iso-temperature surface, \cite{andrews2012, huang2018, qi2024} below which CO is primarily in ices on grains. The intersection of the snow surface with the disk midplane is the CO snow-line, such that CO is entirely released in the gas phase at smaller radii. In the present work, photodissociation is not included, and the impact of gas-grain exchanges is taken into account through depletion factors applied to the ISM gas-phase fractional elemental abundances, as is customary in disk modeling. \cite{andrews2012, huang2018, legal2019, legal2019b, lee2024, cleeves2018} The ISM values are taken from Table~2 of HB18, except for sulfur (see our Table~\ref{tab:models}). Thus, a depletion factor $\depc=8.3\tdix{-5}/\Celem$ is used for the \af{gas-phase} elemental carbon abundance \Celem. Distinct depletion factors characterize the gas phase in the molecular layer and below the snow surface. The oxygen elemental abundance is then determined from the value of \CO\ as $\Oelem = \Celem/\CO$. Although it is expected that the \CO\ should vary spatially, a uniform value \CO=1 is adopted in all the present models. \cite{cleeves2018, lee2024}} \af{Similarly, a depletion factor $\depn=6.4\tdix{-5}/\Nelem$ is employed for the nitrogen gas-phase elemental abundance, but with no distinction above and below the CO snow surface.} Finally, the initial abundances for the gas-phase species in each cell of the disk are not atomic but molecular. They are precalculated using an initially atomic gas with depleted elemental abundances as given in Table~\ref{tab:models}, under physical conditions corresponding to two cells at the outer edge of the disk, respectively above and below the CO snow surface (in practice with $T_\text{gas}\sim 10-30$~K and $\nh\sim 10^{7}$\ccc, see Fig.~\ref{fig:phys_map}).

\begin{figure}[t]
  \includegraphics[height=9cm]{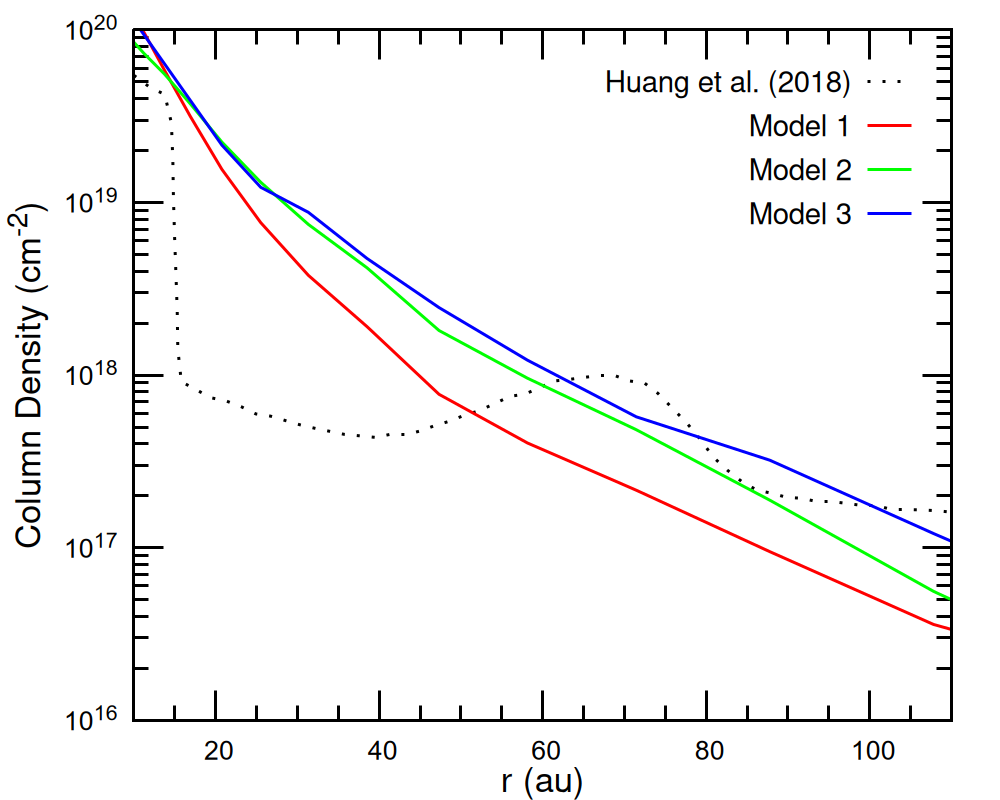}
  \caption{Radial profile of the CO column density for our three models (see Table\,\ref{tab:models}) compared to the ALMA observations of \citet{huang2018}}
  \label{fig:NCO}
\end{figure}

The carbon depletion factors (above and below the snow surface) were constrained by the CO column density measured in TW~Hya \cite{huang2018}. The results are shown in Fig.\,\ref{fig:NCO}. The difference between Models 1 and 2 is due to the exponential tapering of the density profile in the outer disk. It was also found that the vertical temperature profile in Model 3 requires typically ten times larger a depletion below the snow surface in order to reproduce the observed CO column density. In contrast to carbon and oxygen, observational constraints on the nitrogen elemental abundances (or C/N ratios) in disks are very limited due to the lack of dipole-allowed rotational lines of \ce{N2}, the presumably main nitrogen reservoir \cite{cleeves2018}. The nitrogen depletion factor is therefore difficult to quantify, but it could be (loosely) constrained here thanks to the impact of \Nelem\ on the \ce{H2CO} column density, as discussed below. The adopted values of \depc\ and \depn\ are summarized in Table\,\ref{tab:models}. We note that the corresponding N/O ratios are much larger than 1, in line with \citet{cleeves2018}

\section{Results}
\label{sec:results}

The chemistry in each cell of the disk is solved in a time-dependent fashion. Yet, the results presented in this work are steady-state abundances. This simplifying assumption is based on the chemical time scale, $\sim 10$~Myr, being comparable to the age of the TW~Hya system, in the range $7-10$~Myr. \citep{Ruane2017}

The UGAN chemical network being reliable at kinetic temperatures lower than 100~K, our chemical models are limited to the regions of the disk that fulfill this upper limit. In practice, the molecular layer as defined above falls inside this upper limit, except marginally below 90~au in Model~3 (see Fig.\,\ref{fig:phys_map}). Finally, a uniform cosmic-ray (CR) ionization rate was fixed at $\zeta=5\tdix{-19}$ \pers\ \cite{cleeves2015,lee2024} in all models.

According to the temperature structure in our models, the CO snow-line is located at $\approx 15$ au, well within the 11-33 au range derived from ALMA observations. \cite{qi2013,huang2018} In this region, our chemical calculations are not reliable because photodissociation is not included, while the flux from the star (direct and indirect) cannot be neglected. \citep{lee2024} Therefore, we start our calculation at $r=$20~au, corresponding to the observational peak of ortho- and para-\ce{H2CO} column densities. \citep{Terwisscha_van_Scheltinga_2021}

\subsection{Column Density and OPR of Formaldehyde in TW~Hya}
\label{sec:h2co}

\begin{figure}[t]
  \includegraphics[width=0.33\hsize]{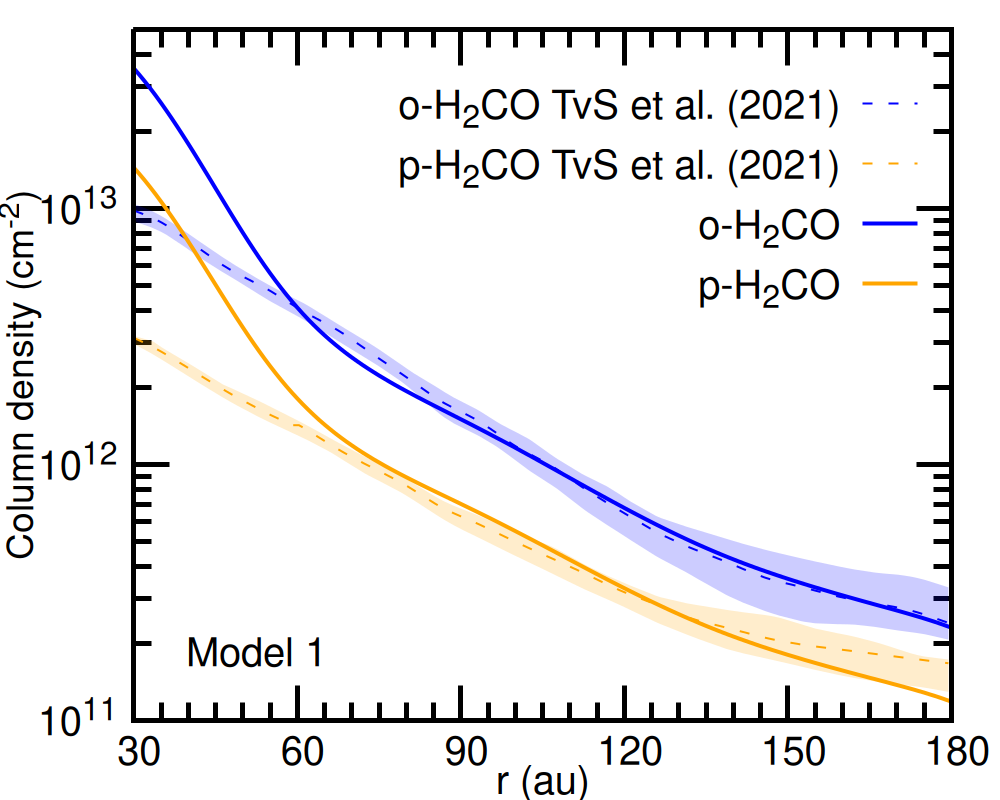}\hfill%
  \includegraphics[width=0.33\hsize]{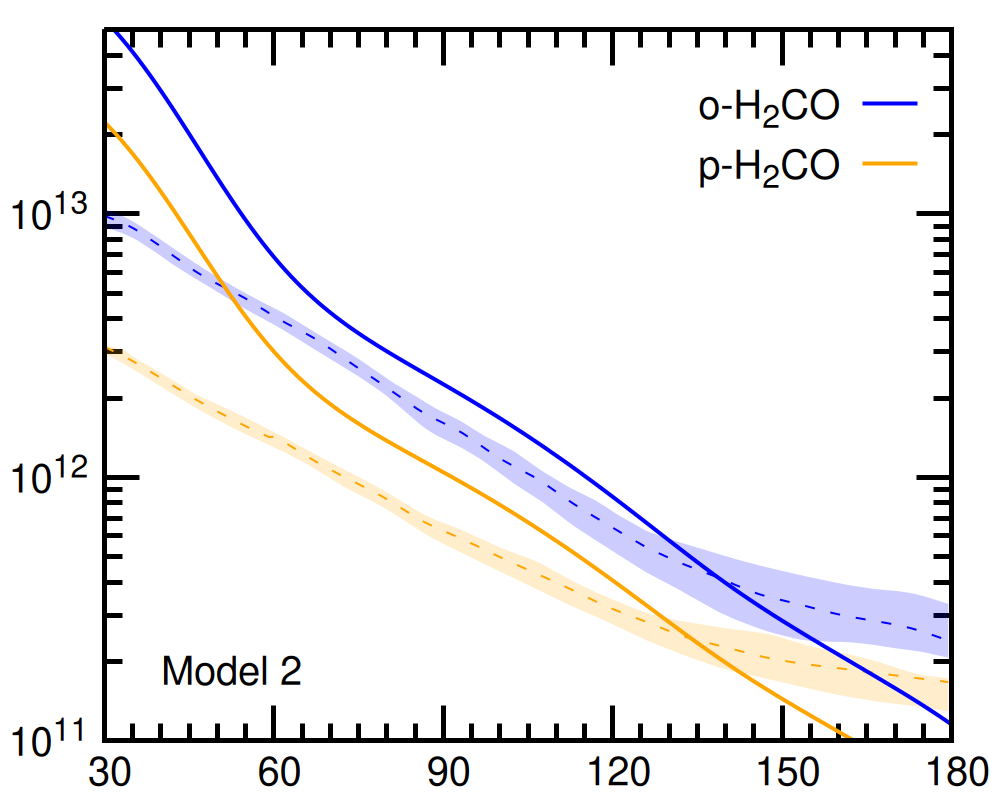}\hfill%
  \includegraphics[width=0.33\hsize]{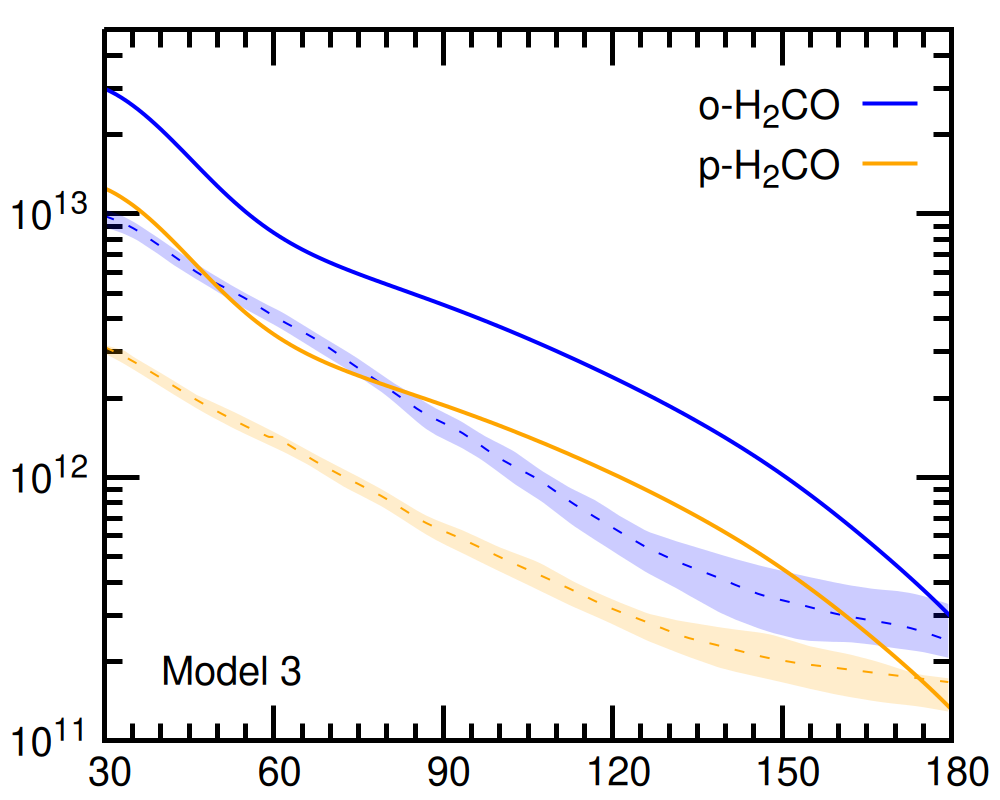}
  \caption{Radial profiles of the column density of ortho- and para-\hhco\ (full lines) for the three models (from left to right). In each panel, the shaded areas show the profiles (with $1\sigma$ uncertainties) derived from Fig.~6 of \citet{Terwisscha_van_Scheltinga_2021} via a digital extraction. The  computed profiles were convolved with a Gaussian beam of 0.5" HPBW (30 au).}
  \label{fig:NH2CO}
\end{figure}

The radial distribution of ortho- and para-\ce{H2CO} provides a critical benchmark for our spin-resolved chemistry model. Figure\,\ref{fig:NH2CO} compares the predicted column densities by our three different models with the ALMA TW~Hya observations. \cite{Terwisscha_van_Scheltinga_2021} We note that all data for the seven detected \ce{H2CO} transitions (3 para and 4 ortho) were brought to the same resolution of 0.5" by \citet{Terwisscha_van_Scheltinga_2021} so that the modeled radial profiles were convolved with a Gaussian kernel of 0.5" HPBW, i.e. 30 au. In doing so, we have artificially extended the value of the \ce{H2CO} column density at $r=$~20~au 
down to $r=4$~au, and set it to zero for lower radii. We checked that the results are only marginally affected if the column density is set to zero for $r < $~20 au.  

Models 1 and 2 reproduce the observed profiles within a factor of 2 between 60 and 150~au, while Model 3 agrees within a factor of 3. The effect of the density tapering is visible in Models 2 and 3 beyond 150 au, while the agreement of Model 1 extends to the outer disk region. The impact of the vertical temperature profile is visible in Model 3 between 60 and 150~au, although the discrepancy remains lower than a factor of three. We note that the nitrogen depletion factor \depn\ was set to 1.5 as a compromise to best reproduce the \ce{H_2CO} column density in all three models. However, no depletion of N ($\depn=0$) cannot be excluded and was indeed found to improve the agreement between observations and Model 3.

\begin{figure}[t]
  \includegraphics[height=8cm]{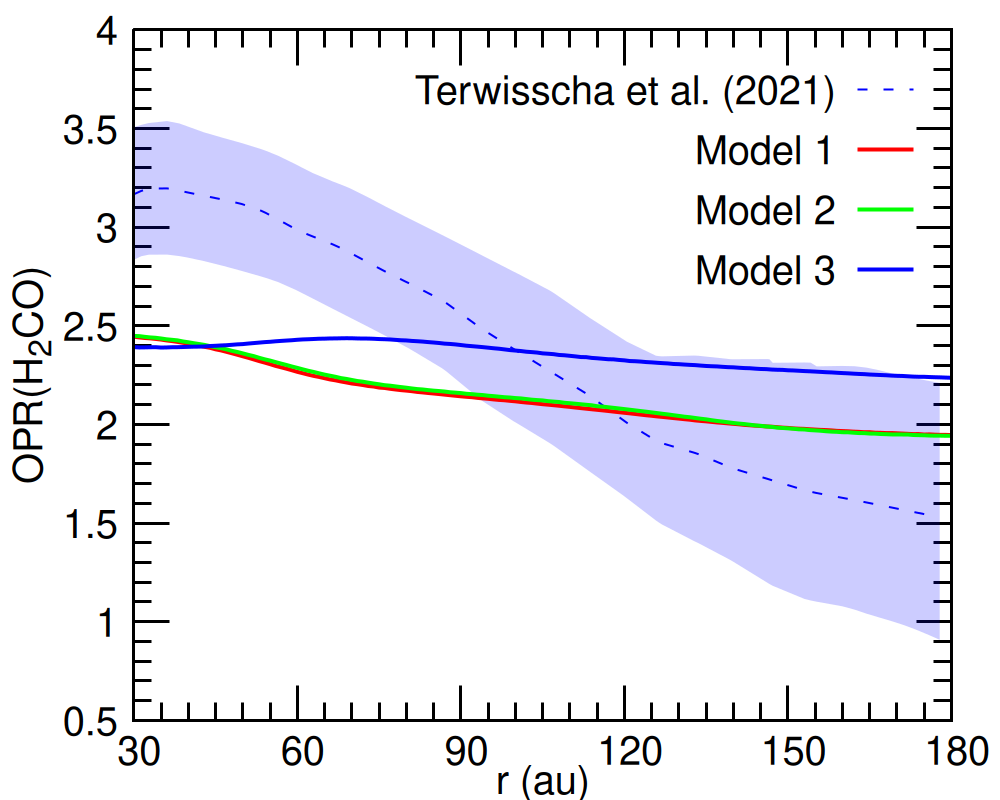}
  \caption{Radial profile of the \hhco\ OPR for the three models. Our results are shown with full lines and the shaded areas show the OPR (with $1\sigma$ uncertainties) derived from ALMA observations. \cite{Terwisscha_van_Scheltinga_2021}}
  \label{fig:OPR}
\end{figure}

Figure\,\ref{fig:OPR} shows the predicted radial profiles of the \ce{H2CO} ortho-to-para ratio. All three models reproduce the observed ratio lower than 2.5 beyond 90~au, with values decreasing down to 1.9 in Models 1 and 2. This strongly suggests that formaldehyde primarily forms in the gas-phase, via the reaction of \ce{CH3} with O (see Section~2.2), in accordance with the nuclear-spin selection rules (derived analytically in the Supporting Information). However, the models predict only weak radial variations of the OPR across the entire disk (beyond 30~au), whereas the observations reveal a strong gradient that rises toward the statistical value of 3 inside $\sim 90$~au. This discrepancy may indicate that additional processes, such as ice thermal- or photodesorption, are involved, especially if radial drift of icy grains is efficient. Indeed, recent laboratory experiments have shown that the gas-phase OPR of sublimated \ce{H2CO} is in the range $\sim 2.9-3$, with no correlation with the temperature of the ice, \cite{yocum2023} thus supporting the idea that the observed inner-disk ratios partly reflect ice-related release and mixing processes.

Among the model parameters, density tapering (Model 2 versus Model 1) has little impact on the OPR, while the vertical temperature profile has a noticeable effect on both the OPR value and its radial variation, as evidenced by the differences between models 2 and 3. This result demonstrates that the OPR of \hhco\ is a sensitive probe of the kinetic temperature in disks, as expected from its dependence on the OPR of \ce{H2} (see Fig.~S2 in Supporting Information), provided that the uncertainties in the derived ortho- and para-\ce{H2CO} column densities are sufficiently small ($\sim 10$\%). 

\subsection{Effects of the C/O Ratio and the CR Ionization Rate}

\begin{figure}[t]
  \centering
  \includegraphics[height=6.6cm]{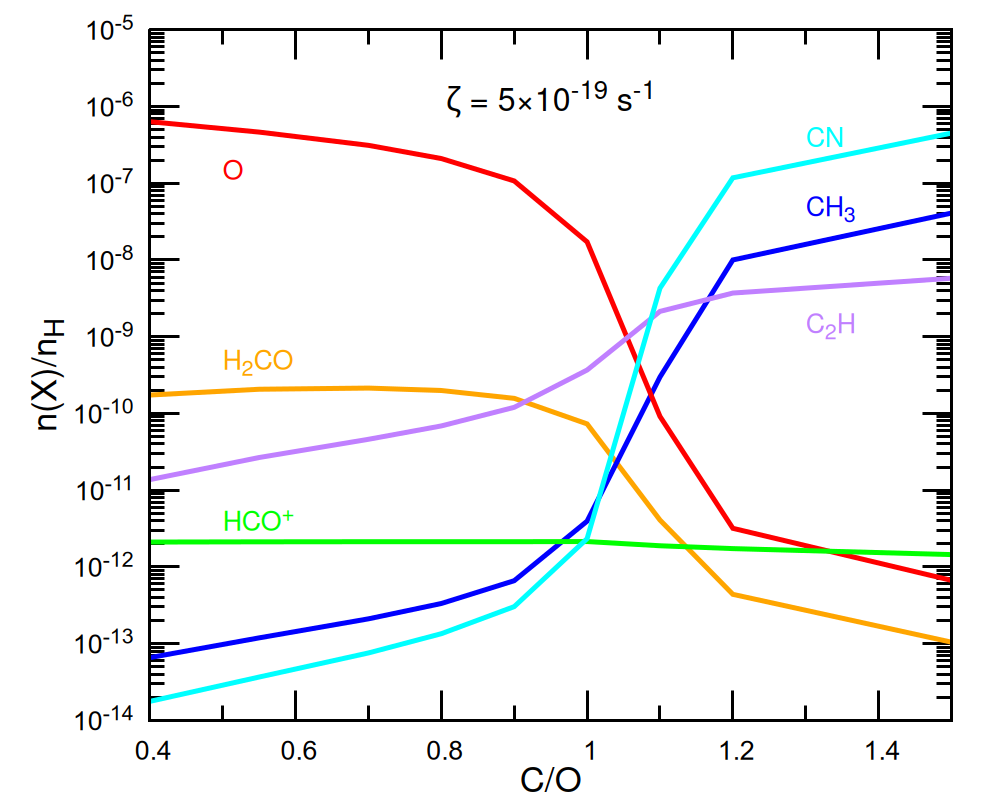}\hfill%
  \includegraphics[height=6.6cm]{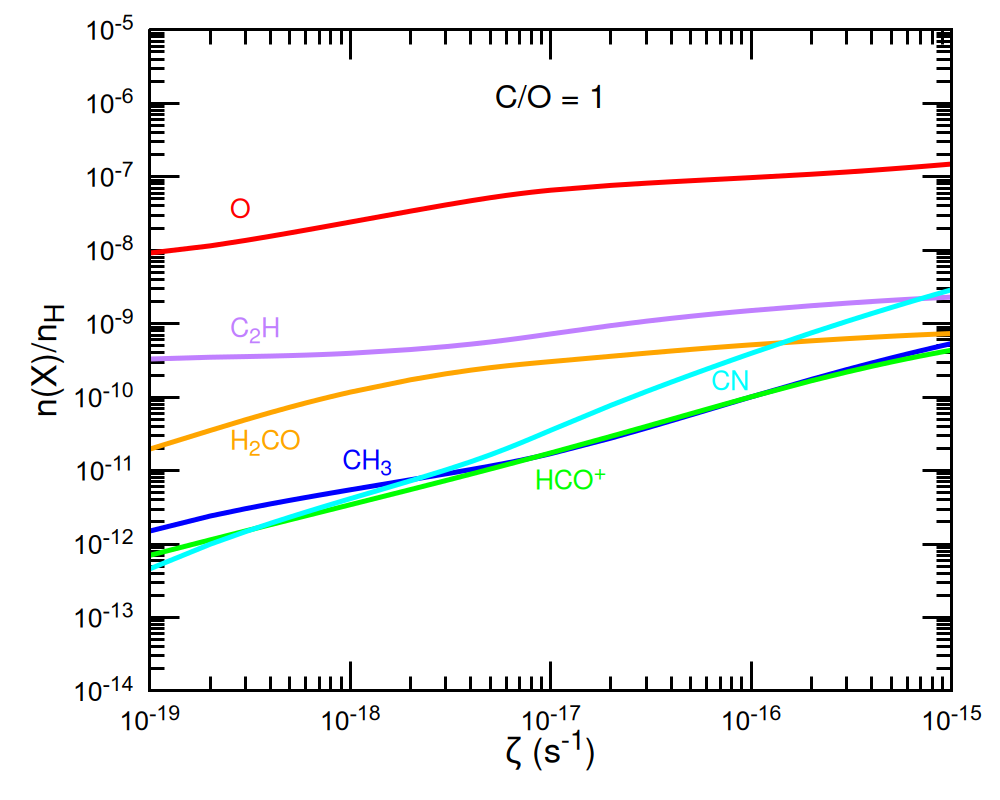}
  \caption{Steady-state abundances of the main precursors and sinks of \hhco, as a function of \CO, for $\zeta=5\tdix{-19}$\pers\ (left) and as a function of $\zeta$ for $\CO=1$ (right). Abundances are computed for a typical cell of the disk models, i.e., $T_\text{gas}=30$K and $\nh=10^{8}$\ccc.}
  \label{fig:param_effect}
\end{figure}

Finally, we investigate the impact of the \CO\ ratio and of $\zeta$ on the abundance and OPR of \hhco\ for typical physical conditions in the disk ($T_\text{gas}=30$~K and $\nh=10^{8}$\ccc, see Fig.~\ref{fig:param_effect}). 

The abundance of \ce{H2CO} is strongly affected by the gas-phase \CO\ ratio. When $\CO \lesssim 1$, sufficient atomic oxygen remains available for the reaction \ce{CH3 + O -> H2CO + H}, and \ce{H2CO} forms efficiently. In contrast, when $\CO \gtrsim 1$, most of the oxygen becomes locked in CO, leaving very little free atomic O in the gas. Despite the concomitant increase of the \ce{CH3} abundance in this regime, the abundance of \ce{H2CO} drops significantly, by up to three orders of magnitude, due to the enhancement of the abundance of \ce{C2H} and CN radicals, which become important \hhco\ destroyers.

The impact of the cosmic-ray ionization rate is qualitatively different. The abundances of \ce{HCO+} and \ce{CH3}, respectively destruction and formation partners of \hhco, both increase with $\zeta$, and the net result is an increase of the \hhco\ abundance. Actually, at $\CO=1$, the destruction of \ce{H2CO} is dominated by \ce{C2H}. 

These results emphasize that the gas-phase abundance of \hhco\ is highly sensitive to both the \CO\ ratio and the cosmic-ray ionization rate. Observations of \hhco\ in multiple disks (where other relevant tracers are detected) could therefore provide valuable constraints on these fundamental but otherwise elusive parameters.

Conversely, the OPR of \ce{H2CO} is found to be largely unaffected by these two parameters (over the investigated ranges), with variations of less than 10\%. This reflects the fact that the OPR of \ce{H2CO} is mostly controlled by the OPR of \ce{H2}, which in turn depends primarily on the kinetic temperature above 15~K (see Supporting Information).

\section{Conclusions and Perspectives}
\label{sec:ccl}

We have presented the first study of the nuclear-spin-resolved chemistry of formaldehyde in disks, focusing on the transition disk TW~Hya. The self-consistent description of nuclear spin states of carbon, nitrogen, oxygen and sulfur hydrides of the UGAN network allows us to compute the spatial distribution of ortho- and para-\ce{H2CO} in a model of TW~Hya that reproduces the CO column density and includes a three-layer disk model (photodissociation, molecular, and freeze-out). Photodissociation by stellar UV and X-ray is not included and the present models focus on the gas-phase chemistry of \ce{H2CO} in the molecular layer. The present physicochemical disk model reproduces the observed radial profile of the ortho- and para-\ce{H2CO} column density to within a factor of two from 60 to 150 au. The derived ortho-to-para ratio of \ce{H2CO} is also well reproduced, especially the values lower than the statistical ratio of 3 measured in this disk. In contrast with the observations, our model predicts a relatively constant radial profile, with a ratio of $\sim 1.9-2.5$, which is lower than observed at radii below 70~au. It is found that the vertical temperature profile has a noticeable effect on the OPR of \ce{H2CO}. However, the uncertainties in the observed OPR value and radial profile are currently too large to make this OPR a sensitive probe of the vertical temperature profile. Future observations including more transitions of \ce{H2CO} would be highly valuable. The discrepancy between the predicted and measured OPR at radii below 70 au suggests that processes such as desorption combined with vertical/radial mixing could be important in this region, leading to higher values of the OPR than obtained by our present model. The total abundance of \ce{H2CO} is also found to be highly sensitive to the elemental ratio \CO\ in the gas-phase, decreasing by three orders of magnitude at C/O $>1.2$. Similarly, the cosmic-ray ionization rate also has a strong impact on the abundance of \ce{H2CO}.
Observations of \ce{H2CO} in a sample of disks would provide an observational test of the dependence of its abundance on \CO\ and $\zeta$.

\begin{suppinfo}
  \begin{itemize}
  \item Figure S1. Details of the temperature and density profiles.
  \item Figure S2. Analytical ortho-to-para ratio of \ce{H2CO}.
  \end{itemize}
\end{suppinfo}

\begin{acknowledgement}
We thank the anonymous reviewers for their careful reading and valuable comments. The authors acknowledge the financial support of the University Grenoble Alpes for funding Hideko Nomura as a Visiting Professor. This research was also supported by the International Space Science Institute (ISSI) in Bern, through ISSI International Team project ``Understanding Nuclear Spin Temperatures in Astronomical Environments ISSI Team 25-648''. RLG also thanks the financial support from the French Agence Nationale de la
Recherche (ANR) through the project MAPSAJE (ANR-24-CE31-2126-01).
\end{acknowledgement}

\bibliography{refs}

@ARTICLE{guzman2018,
       author = {{Guzm{\'a}n}, V.~V. and {{\"O}berg}, K.~I. and {Carpenter}, J. and {Le Gal}, R. and {Qi}, C. and {Pagues}, J.},
        title = "{H$_{2}$CO Ortho-to-para Ratio in the Protoplanetary Disk HD 163296}",
      journal = {Astrophys. J.},
         year = 2018,
        month = sep,
       volume = {864},
       number = {2},
          eid = {170},
        pages = {170},
          doi = {10.3847/1538-4357/aad778},
archivePrefix = {arXiv},
       eprint = {1809.01705},
 primaryClass = {astro-ph.GA},
       adsurl = {https://ui.adsabs.harvard.edu/abs/2018ApJ...864..170G}
}

@ARTICLE{yocum2023,
       author = {{Yocum}, K.~M. and {Wilkins}, O.~H. and {Bardwell}, J.~C. and {Milam}, S.~N. and {Gerakines},
P.~A.},
        title = "{Gas-phase Ortho-to-para Ratio of Formaldehyde Formed at Low Temperatures in Laboratory
Ices}",
      journal = {Astrophys. J. Lett.},
         year = 2023,
        month = dec,
       volume = {958},
       number = {2},
          eid = {L41},
        pages = {L41},
          doi = {10.3847/2041-8213/ad0bee},
       adsurl = {https://ui.adsabs.harvard.edu/abs/2023ApJ...958L..41Y}
}

@Article{dartois2003,
  Title                    = {{Structure of the DM Tau Outer Disk: Probing the vertical kinetic temperature gradient}},
  Author                   = {{Dartois}, E. and {Dutrey}, A. and {Guilloteau}, S.},
  Journal                  = {Astron. Astrophys.},
  Year                     = {2003},

  Month                    = {feb},
  Pages                    = {773-787},
  Volume                   = {399}
}

@ARTICLE{Faure2013,
       author = {{Faure}, A. and {Hily-Blant}, P. and {Le Gal}, R. and {Rist}, C. and {Pineau des For{\^e}ts}, G.},
        title = "{Ortho-Para Selection Rules in the Gas-phase Chemistry of Interstellar Ammonia}",
      journal = {Astrophys. J.},
         year = 2013,
        month = jun,
       volume = {770},
       number = {1},
          eid = {L2},
        pages = {L2},
          doi = {10.1088/2041-8205/770/1/L2},
       adsurl = {https://ui.adsabs.harvard.edu/abs/2013ApJ...770L...2F}
}

@article{hama_ortho--para_2018,
	title = {The {Ortho}-to-para {Ratio} of {Water} {Molecules} {Desorbed} from {Ice} {Made} from {Para}-water {Monomers} at 11 {K}},
	volume = {857},
	issn = {2041-8205, 2041-8213},
	url = {https://iopscience.iop.org/article/10.3847/2041-8213/aabc0c},
	doi = {10.3847/2041-8213/aabc0c},
	number = {2},
	journal = {Astrophys. J. Lett.},
	author = {Hama, Tetsuya and Kouchi, Akira and Watanabe, Naoki},
	month = apr,
	year = {2018},
	pages = {L13},
}

@article{mumma_chemical_2011,
	title = {The {Chemical} {Composition} of {Comets}—{Emerging} {Taxonomies} and {Natal} {Heritage}},
	volume = {49},
	issn = {0066-4146, 1545-4282},
	url = {https://www.annualreviews.org/doi/10.1146/annurev-astro-081309-130811},
	doi = {10.1146/annurev-astro-081309-130811},
	language = {en},
	number = {1},
	journal = {Annu. Rev. Astron. Astrophys.},
	author = {Mumma, Michael J. and Charnley, Steven B.},
	month = sep,
	year = {2011},
	pages = {471--524},
}

@ARTICLE{quack1977,
       author = {{Quack}, Martin},
        title = "{Detailed symmetry selection rules for reactive collisions}",
      journal = {Mol. Phys.},
         year = 1977,
        month = jan,
       volume = {34},
       number = {2},
        pages = {477-504},
          doi = {10.1080/00268977700101861},
       adsurl = {https://ui.adsabs.harvard.edu/abs/1977MolPh..34..477Q}
}

@ARTICLE{legal2019,
       author = {{Le Gal}, Romane and {{\"O}berg}, Karin I. and {Loomis}, Ryan A. and {Pegues}, Jamila and {Bergner}, Jennifer B.},
        title = "{Sulfur Chemistry in Protoplanetary Disks: CS and H$_{2}$CS}",
      journal = {Astrophys. J.},
         year = 2019,
        month = may,
       volume = {876},
       number = {1},
          eid = {72},
        pages = {72},
          doi = {10.3847/1538-4357/ab1416},
archivePrefix = {arXiv},
       eprint = {1903.11105},
 primaryClass = {astro-ph.GA},
       adsurl = {https://ui.adsabs.harvard.edu/abs/2019ApJ...876...72L}
}

@ARTICLE{legal2019b,
       author = {{Le Gal}, Romane and {Brady}, Madison T. and {{\"O}berg}, Karin I. and {Roueff}, Evelyne and {Le Petit}, Franck},
        title = "{The Role of C/O in Nitrile Astrochemistry in PDRs and Planet-forming Disks}",
      journal = {Astrophys. J.},
         year = 2019,
        month = dec,
       volume = {886},
       number = {2},
          eid = {86},
        pages = {86},
          doi = {10.3847/1538-4357/ab4ad9},
archivePrefix = {arXiv},
       eprint = {1910.01554},
 primaryClass = {astro-ph.EP},
       adsurl = {https://ui.adsabs.harvard.edu/abs/2019ApJ...886...86L}
}

@ARTICLE{legal2017,
       author = {{Le Gal}, R. and {Xie}, C. and {Herbst}, E. and {Talbi}, D. and {Guo}, H. and {Muller}, S.},
        title = "{The ortho-to-para ratio of H$_{2}$Cl$^{+}$: Quasi-classical trajectory calculations and new simulations in light of new observations}",
      journal = {Astron. Astrophys.},
         year = 2017,
        month = dec,
       volume = {608},
          eid = {A96},
        pages = {A96},
          doi = {10.1051/0004-6361/201731566},
archivePrefix = {arXiv},
       eprint = {1708.08980},
 primaryClass = {astro-ph.GA},
       adsurl = {https://ui.adsabs.harvard.edu/abs/2017A&A...608A..96L}
}

@ARTICLE{lee2021,
       author = {{Lee}, Seokho and {Nomura}, Hideko and {Furuya}, Kenji and {Lee}, Jeong-Eun},
        title = "{Modeling Nitrogen Fractionation in the Protoplanetary Disk around TW Hya: Model Constraints on Grain Population and Carbon-to-oxygen Elemental Abundance Ratio}",
      journal = {Astrophys. J.},
         year = 2021,
        month = feb,
       volume = {908},
       number = {1},
          eid = {82},
        pages = {82},
          doi = {10.3847/1538-4357/abd633},
archivePrefix = {arXiv},
       eprint = {2012.12531},
 primaryClass = {astro-ph.GA},
       adsurl = {https://ui.adsabs.harvard.edu/abs/2021ApJ...908...82L}
}

@article{lee2024,
	title = {Carbon {Isotope} {Chemistry} in {Protoplanetary} {Disks}: {Effects} of {C}/{O} {Ratios}},
	volume = {969},
	issn = {0004-637X, 1538-4357},
	shorttitle = {Carbon {Isotope} {Chemistry} in {Protoplanetary} {Disks}},
	url = {https://iopscience.iop.org/article/10.3847/1538-4357/ad39e3},
	doi = {10.3847/1538-4357/ad39e3},
	number = {1},
	urldate = {2025-02-12},
	journal = {Astrophys. J.},
	author = {Lee, Seokho and Nomura, Hideko and Furuya, Kenji},
	month = jul,
	year = {2024},
	pages = {41},
}

@article{kama2016,
	author = {{Kama}, M. and {Bruderer, S.} and {van Dishoeck, E. F.} and {Hogerheijde, M.} and {Folsom, C. P.} and {Miotello, A.} and {Fedele, D.} and {Belloche, A.} and {Güsten, R.} and {Wyrowski, F.}},
	title = {Volatile-carbon locking and release in protoplanetary disks - A study of TW Hya and HD 100546},
	DOI= "10.1051/0004-6361/201526991",
	url= "https://doi.org/10.1051/0004-6361/201526991",
	journal = {Astron. Astrophys.},
	year = 2016,
	volume = 592,
	pages = "A83",
}

@article{huang2018,
doi = {10.3847/1538-4357/aaa1e7},
url = {https://dx.doi.org/10.3847/1538-4357/aaa1e7},
year = {2018},
month = {jan},
publisher = {The American Astronomical Society},
volume = {852},
number = {2},
pages = {122},
author = {Huang, Jane and Andrews, Sean M. and Cleeves, L. Ilsedore and Öberg, Karin I. and Wilner, David J. and Bai, Xuening and Birnstiel, Til and Carpenter, John and Hughes, A. Meredith and Isella, Andrea and Pérez, Laura M. and Ricci, Luca and Zhu, Zhaohuan},
title = {CO and Dust Properties in the TW Hya Disk from High-resolution ALMA Observations},
journal = {Astrophys. J.},
}

@article{cleeves2015,
doi = {10.1088/0004-637X/799/2/204},
url = {https://dx.doi.org/10.1088/0004-637X/799/2/204},
year = {2015},
month = {jan},
publisher = {The American Astronomical Society},
volume = {799},
number = {2},
pages = {204},
author = {L. Ilsedore Cleeves and Edwin A. Bergin and Chunhua Qi and Fred C. Adams and Karin I. Öberg},
title = {Constraining the x-ray and cosmic-ray ionization chemistry of the TW Hya protoplanetary disk: evidence for a sub-interstellar cosmic-ray rate},
journal = {Astrophys. J.},
}

@article{rosenfeld2013,
doi = {10.1088/0004-637X/775/2/136},
url = {https://dx.doi.org/10.1088/0004-637X/775/2/136},
year = {2013},
month = {sep},
publisher = {The American Astronomical Society},
volume = {775},
number = {2},
pages = {136},
author = {Katherine A. Rosenfeld and Sean M. Andrews and David J. Wilner and J. H. Kastner and M. K. McClure},
title = {The structure of the evolved circumbinary disk around V4046 Sgr},
journal = {Astrophys. J.},
}

@article{andrews2012,
doi = {10.1088/0004-637X/744/2/162},
url = {https://dx.doi.org/10.1088/0004-637X/744/2/162},
year = {2011},
month = {dec},
publisher = {The American Astronomical Society},
volume = {744},
number = {2},
pages = {162},
author = {Sean M. Andrews and David J. Wilner and A. M. Hughes and Chunhua Qi and Katherine A. Rosenfeld and Karin I. Öberg and T. Birnstiel and Catherine Espaillat and Lucas A. Cieza and Jonathan P. Williams and Shin-Yi Lin and Paul T. P. Ho},
title = {The TW Hya disk at 870mic: comparison of CO and dust radial structures},
journal = {Astrophys. J.},
}

@ARTICLE{wagenblast1989,
       author = {{Wagenblast}, R. and {Hartquist}, T.~W.},
        title = "{Non-equilibrium level populations of molecular hydrogen -II. Models of ZET OPH cloud.}",
      journal = {Mon. Not. R. Astron. Soc.},
         year = 1989,
        month = apr,
       volume = {237},
        pages = {1019-1025},
          doi = {10.1093/mnras/237.4.1019},
       adsurl = {https://ui.adsabs.harvard.edu/abs/1989MNRAS.237.1019W}
}

@article{canta2021,
doi = {10.3847/1538-4357/ac23da},
url = {https://dx.doi.org/10.3847/1538-4357/ac23da},
year = {2021},
month = {nov},
publisher = {The American Astronomical Society},
volume = {922},
number = {1},
pages = {62},
author = {Alessandra Canta and Richard Teague and Romane Le Gal and Karin I. Öberg},
title = {The First Detection of CH2CN in a Protoplanetary Disk},
journal = {Astrophys. J.}
}

@ARTICLE{HB2018,
       author = {{Hily-Blant}, P. and {Faure}, A. and {Rist}, C. and {Pineau des For{\^e}ts}, G. and {Flower}, D.~R.},
        title = "{Modelling the molecular composition and nuclear-spin chemistryof collapsing pre-stellar sources}",
      journal = {Mon. Not. R. Astron. Soc.},
         year = 2018,
        month = jul,
       volume = {477},
       number = {4},
        pages = {4454-4472},
          doi = {10.1093/mnras/sty881},
archivePrefix = {arXiv},
       eprint = {1804.01354},
 primaryClass = {astro-ph.GA},
       adsurl = {https://ui.adsabs.harvard.edu/abs/2018MNRAS.477.4454H}
}

@ARTICLE{oka2004,
       author = {{Oka}, Takeshi},
        title = "{Nuclear spin selection rules in chemical reactions by angular momentum algebra}",
      journal = {Journal of Molecular Spectroscopy},
         year = 2004,
        month = dec,
       volume = {228},
       number = {2},
        pages = {635-639},
          doi = {10.1016/j.jms.2004.08.015},
       adsurl = {https://ui.adsabs.harvard.edu/abs/2004JMoSp.228..635O}
}

@article{Terwisscha_van_Scheltinga_2021,
doi = {10.3847/1538-4357/abc9ba},
url = {https://dx.doi.org/10.3847/1538-4357/abc9ba},
year = {2021},
month = {jan},
publisher = {The American Astronomical Society},
volume = {906},
number = {2},
pages = {111},
author = {{Terwisscha van Scheltinga}, J. and Michiel R. Hogerheijde and L. Ilsedore Cleeves and Ryan A. Loomis and Catherine Walsh and Karin I. Öberg and Edwin A. Bergin and Jennifer B. Bergner and Geoffrey A. Blake and Jenny K. Calahan and Paolo Cazzoletti and Ewine F. van Dishoeck and Viviana V. Guzmán and Jane Huang and Mihkel Kama and Chunhua Qi and Richard Teague and David J. Wilner},
title = {The TW Hya Rosetta Stone Project. II. Spatially Resolved Emission of Formaldehyde Hints at Low-temperature Gas-phase Formation},
journal = {Astrophys. J.}
}

@ARTICLE{Henning2013,
       author = {{Henning}, Thomas and {Semenov}, Dmitry},
        title = "{Chemistry in Protoplanetary Disks}",
      journal = {Chem. Rev.},
         year = 2013,
        month = dec,
       volume = {113},
       number = {12},
        pages = {9016-9042},
          doi = {10.1021/cr400128p},
archivePrefix = {arXiv},
       eprint = {1310.3151},
 primaryClass = {astro-ph.GA},
       adsurl = {https://ui.adsabs.harvard.edu/abs/2013ChRv..113.9016H}
}

@ARTICLE{Faure2019,
       author = {{Faure}, A. and {Hily-Blant}, P. and {Rist}, C. and {Pineau des For{\^e}ts}, G. and {Matthews}, A. and {Flower}, D.~R.},
        title = "{The ortho-to-para ratio of water in interstellar clouds}",
      journal = {Mon. Not. R. Astron. Soc.},
         year = 2019,
        month = aug,
       volume = {487},
       number = {3},
        pages = {3392-3403},
          doi = {10.1093/mnras/stz1531},
archivePrefix = {arXiv},
       eprint = {2201.02068},
 primaryClass = {astro-ph.GA},
       adsurl = {https://ui.adsabs.harvard.edu/abs/2019MNRAS.487.3392F}
}

@article{Hu2021,
  title={Nuclear spin conservation enables state-to-state control of ultracold molecular reactions},
  author={Hu, Ming-Guang and Liu, Yu and Nichols, Matthew A and Zhu, Lingbang and Qu{\'e}m{\'e}ner, Goulven and Dulieu, Olivier and Ni, Kang-Kuen},
  journal={Nat. Chem.},
  volume={13},
  number={5},
  pages={435--440},
  year={2021},
  publisher={Nature Publishing Group UK London}
}

@ARTICLE{qi2013,
       author = {{Qi}, Chunhua and {{\"O}berg}, Karin I. and {Wilner}, David J. and {D'Alessio}, Paola and {Bergin}, Edwin and {Andrews}, Sean M. and {Blake}, Geoffrey A. and {Hogerheijde}, Michiel R. and {van Dishoeck}, Ewine F.},
        title = "{Imaging of the CO Snow Line in a Solar Nebula Analog}",
      journal = {Science},
         year = 2013,
        month = aug,
       volume = {341},
       number = {6146},
        pages = {630-632},
          doi = {10.1126/science.1239560},
archivePrefix = {arXiv},
       eprint = {1307.7439},
 primaryClass = {astro-ph.SR},
       adsurl = {https://ui.adsabs.harvard.edu/abs/2013Sci...341..630Q}
}

@ARTICLE{qi2024,
       author = {{Qi}, Chunhua and {Wilner}, David J.},
        title = "{Evidence for a Sharp CO Snow Line Transition in a Protoplanetary Disk and Implications for Millimeter-wave Observations of CO Isotopologues}",
      journal = {Astrophys. J.},
         year = 2024,
        month = dec,
       volume = {977},
       number = {1},
          eid = {60},
        pages = {60},
          doi = {10.3847/1538-4357/ad8d55},
archivePrefix = {arXiv},
       eprint = {2410.23036},
 primaryClass = {astro-ph.EP},
       adsurl = {https://ui.adsabs.harvard.edu/abs/2024ApJ...977...60Q}
}

@article{cleeves2018,
	title = {Constraining {Gas}-phase {Carbon}, {Oxygen}, and {Nitrogen} in the {IM} {Lup} {Protoplanetary} {Disk}},
	volume = {865},
	issn = {0004-637X, 1538-4357},
	url = {https://iopscience.iop.org/article/10.3847/1538-4357/aade96},
	doi = {10.3847/1538-4357/aade96},
	number = {2},
	journal = {Astrophys. J.},
	author = {Cleeves, L. and Öberg, Karin I. and Wilner, David J. and Huang, Jane and Loomis, Ryan A. and Andrews, Sean M. and Guzman, V. V.},
	month = oct,
	year = {2018},
	pages = {155},
}

@article{Klemperer2010,
   author = "Klemperer, William",
   title = "Astronomical Chemistry", 
   journal= "Annu. Rev. Phys. Chem.",
   year = "2011",
   volume = "62",
   number = "Volume 62, 2011",
   pages = "173-184",
   doi = "https://doi.org/10.1146/annurev-physchem-032210-103332",
   url = "https://www.annualreviews.org/content/journals/10.1146/annurev-physchem-032210-103332",
   publisher = "Annual Reviews",
   issn = "1545-1593",
   type = "Journal Article",
  }

@article{Xu2015,
author = {Xu, Z. F. and Raghunath, P. and Lin, M. C.},
title = {Ab Initio Chemical Kinetics for the CH3 + O(3P) Reaction and Related Isomerization–Decomposition of CH3O and CH2OH Radicals},
journal = {J. Phys. Chem. A},
volume = {119},
number = {28},
pages = {7404-7417},
year = {2015},
doi = {10.1021/acs.jpca.5b00553},
    note ={PMID: 25751420},
URL = {https://doi.org/10.1021/acs.jpca.5b00553},
eprint = {https://doi.org/10.1021/acs.jpca.5b00553}
}

@Article{Hack2005,
author ="Hack, W. and Hold, M. and Hoyermann, K. and Wehmeyer, J. and Zeuch, T.",
title  ="Mechanism and rate of the reaction CH3 + O—revisited",
journal  ="Phys. Chem. Chem. Phys.",
year  ="2005",
volume  ="7",
issue  ="9",
pages  ="1977-1984",
publisher  ="The Royal Society of Chemistry",
doi  ="10.1039/B419137D",
url  ="http://dx.doi.org/10.1039/B419137D",
}

@ARTICLE{HB2020,
       author = {{Hily-Blant}, P. and {Pineau des For{\^e}ts}, G. and {Faure}, A. and {Flower}, D.~R.},
        title = "{Depletion and fractionation of nitrogen in collapsing cores}",
      journal = {Astron. Astrophys.},
         year = 2020,
        month = nov,
       volume = {643},
          eid = {A76},
        pages = {A76},
          doi = {10.1051/0004-6361/202038780},
archivePrefix = {arXiv},
       eprint = {2009.06393},
 primaryClass = {astro-ph.GA},
       adsurl = {https://ui.adsabs.harvard.edu/abs/2020A\&A...643A..76H}
}

@ARTICLE{HB2022,
       author = {{Hily-Blant}, P. and {Pineau des For{\^e}ts}, G. and {Faure}, A. and {Lique}, F.},
        title = "{Sulfur gas-phase abundance in dense cores}",
      journal = {Astron. Astrophys.},
         year = 2022,
        month = feb,
       volume = {658},
          eid = {A168},
        pages = {A168},
          doi = {10.1051/0004-6361/201936498},
archivePrefix = {arXiv},
       eprint = {2112.01076},
 primaryClass = {astro-ph.GA},
       adsurl = {https://ui.adsabs.harvard.edu/abs/2022A\&A...658A.168H}
}

@ARTICLE{Legal2014,
       author = {{Le Gal}, R. and {Hily-Blant}, P. and {Faure}, A. and {Pineau des For{\^e}ts}, G. and {Rist}, C. and {Maret}, S.},
        title = "{Interstellar chemistry of nitrogen hydrides in dark clouds}",
      journal = {Astron. Astrophys.},
         year = 2014,
        month = feb,
       volume = {562},
          eid = {A83},
        pages = {A83},
          doi = {10.1051/0004-6361/201322386},
archivePrefix = {arXiv},
       eprint = {1311.5313},
 primaryClass = {astro-ph.SR},
       adsurl = {https://ui.adsabs.harvard.edu/abs/2014A\&A...562A..83L}
}

@ARTICLE{Watanabe2002,
       author = {{Watanabe}, Naoki and {Kouchi}, Akira},
        title = "{Efficient Formation of Formaldehyde and Methanol by the Addition of Hydrogen Atoms to CO in H$_{2}$O-CO Ice at 10 K}",
      journal = {Astrophys. J. Lett.},
         year = 2002,
        month = jun,
       volume = {571},
       number = {2},
        pages = {L173-L176},
          doi = {10.1086/341412},
       adsurl = {https://ui.adsabs.harvard.edu/abs/2002ApJ...571L.173W}
}

@ARTICLE{Faure2025,
       author = {{Faure}, M. and {Bacmann}, A. and {Faure}, A. and {Quirico}, E. and {Boduch}, P. and {Domaracka}, A. and {Rothard}, H.},
        title = "{Cosmic-ray induced sputtering of interstellar formaldehyde ices}",
      journal = {Astron. Astrophys.},
         year = 2025,
        month = jan,
       volume = {693},
          eid = {A30},
        pages = {A30},
          doi = {10.1051/0004-6361/202449937},
archivePrefix = {arXiv},
       eprint = {2409.01700},
 primaryClass = {astro-ph.GA},
       adsurl = {https://ui.adsabs.harvard.edu/abs/2025A\&A...693A..30F}
}

@ARTICLE{Su1982,
       author = {{Su}, Timothy and {Chesnavich}, Walter J.},
        title = "{Parametrization of the ion-polar molecule collision rate constant by trajectory calculations}",
      journal = {J. Chem. Phys.},
         year = 1982,
        month = may,
       volume = {76},
       number = {10},
        pages = {5183-5185},
          doi = {10.1063/1.442828},
       adsurl = {https://ui.adsabs.harvard.edu/abs/1982JChPh..76.5183S}
}

@ARTICLE{Douglas2024,
       author = {{Douglas}, Kevin M. and {West}, Niclas A. and {Lucas}, Daniel I. and {Van de Sande}, Marie and {Blitz}, Mark A. and {Heard}, Dwayne E.},
        title = "{Experimental and Theoretical Investigation of the Reaction of C2H with Formaldehyde (CH2O) at Very Low Temperatures and Application to Astrochemical Models}",
      journal = {ACS Earth Space Chem.},
         year = 2024,
        month = dec,
       volume = {8},
       number = {12},
        pages = {2428-2441},
          doi = {10.1021/acsearthspacechem.4c00188},
       adsurl = {https://ui.adsabs.harvard.edu/abs/2024ESC.....8.2428D}
}

@ARTICLE{West2023,
       author = {{West}, Niclas A. and {Li}, Lok Hin Desmond and {Millar}, Tom J. and {Van de Sande}, Marie and {Rutter}, Edward and {Blitz}, Mark A. and {Lehman}, Julia H. and {Decin}, Leen and {Heard}, Dwayne E.},
        title = "{Experimental and theoretical study of the low-temperature kinetics of the reaction of CN with CH2O and implications for interstellar environments}",
      journal = {Phys. Chem. Chem. Phys.},
         year = 2023,
        month = mar,
       volume = {25},
       number = {11},
        pages = {7719-7733},
          doi = {10.1039/D2CP05043A},
       adsurl = {https://ui.adsabs.harvard.edu/abs/2023PCCP...25.7719W}
}

@article{Heard2018,
author = {Heard, Dwayne E.},
title = {Rapid Acceleration of Hydrogen Atom Abstraction Reactions of OH at Very Low Temperatures through Weakly Bound Complexes and Tunneling},
journal = {Acc. Chem. Res.},
volume = {51},
number = {11},
pages = {2620-2627},
year = {2018},
doi = {10.1021/acs.accounts.8b00304},
URL = {https://doi.org/10.1021/acs.accounts.8b00304},
eprint = {https://doi.org/10.1021/acs.accounts.8b00304}
}

@ARTICLE{Hillenbrand2022,
       author = {{Hillenbrand}, Pierre-Michel and {de Ruette}, Nathalie and {Urbain}, Xavier and {Savin}, Daniel W.},
        title = "{Branching Ratio for O + H$_{3}$ $^{+}$ Forming OH$^{+}$ + H$_{2}$ and H$_{2}$O$^{+}$ + H}",
      journal = {Astrophys. J.},
         year = 2022,
        month = mar,
       volume = {927},
       number = {1},
          eid = {47},
        pages = {47},
          doi = {10.3847/1538-4357/ac41ce},
       adsurl = {https://ui.adsabs.harvard.edu/abs/2022ApJ...927...47H}
}

@ARTICLE{Kaminska2010,
       author = {{Kami{\'n}ska}, Magdalena and {Zhaunerchyk}, Vitali and {Vigren}, Erik and {Danielsson}, Mathias and {Hamberg}, Mathias and {Geppert}, Wolf D. and {Larsson}, Mats and {Ros{\'e}n}, Stefan and {Thomas}, Richard D. and {Semaniak}, Jacek},
        title = "{Dissociative recombination of CH$_{5}$$^{+}$ and CD$_{5}$$^{+}$: Measurement of the product branching fractions and the absolute cross sections, and the breakup dynamics in the CH$_{3}$+H+H product channel}",
      journal = {Phys. Rev. A},
         year = 2010,
        month = jun,
       volume = {81},
       number = {6},
          eid = {062701},
        pages = {062701},
          doi = {10.1103/PhysRevA.81.062701},
       adsurl = {https://ui.adsabs.harvard.edu/abs/2010PhRvA..81f2701K}
}

@ARTICLE{Gerlich2005,
       author = {{Gerlich}, D.},
        title = "{Probing the structure of CH5+ ions and deuterated variants via collisions}",
      journal = {Phys. Chem. Chem. Phys.},
         year = 2005,
        month = jan,
       volume = {7},
       number = {7},
        pages = {1583},
          doi = {10.1039/B419328H},
       adsurl = {https://ui.adsabs.harvard.edu/abs/2005PCCP....7.1583G}
}

@ARTICLE{Rist2013,
       author = {{Rist}, Claire and {Faure}, Alexandre and {Hily-Blant}, Pierre and {Le Gal}, Romane},
        title = "{Nuclear-Spin Selection Rules in the Chemistry of Interstellar Nitrogen Hydrides}",
      journal = {J. Phys. Chem. A},
         year = 2013,
        month = oct,
       volume = {117},
       number = {39},
        pages = {9800-9806},
          doi = {10.1021/jp312640a},
       adsurl = {https://ui.adsabs.harvard.edu/abs/2013JPCA..117.9800R}
}

@ARTICLE{Hamberg2007,
       author = {{Hamberg}, M. and {Geppert}, W.~D. and {Thomas}, R.~D. and {Zhaunerchyk}, V. and {{\"O}sterdahl}, F. and {Ehlerding}, A. and {Kaminska}, M. and {Semaniak}, J. and {Ugglas}, M. Af and {K{\"a}llberg}, A. and {Paal}, A. and {Simonsson}, A. and {Larsson}, M.},
        title = "{Experimental determination of dissociative recombination reaction pathways and absolute reaction cross-sections of CH2OH+, CD2OD+ and CD2}",
      journal = {Molecular Physics},
         year = 2007,
        month = mar,
       volume = {105},
       number = {5},
        pages = {899-906},
          doi = {10.1080/00268970701206642},
       adsurl = {https://ui.adsabs.harvard.edu/abs/2007MolPh.105..899H}
}

@ARTICLE{Flower2006a,
       author = {{Flower}, D.~R. and {Pineau Des For{\^e}ts}, G. and {Walmsley}, C.~M.},
        title = "{The abundances of nitrogen-containing molecules during pre-protostellar collapse}",
      journal = {Astron. Astrophys.},
         year = 2006,
        month = sep,
       volume = {456},
       number = {1},
        pages = {215-223},
          doi = {10.1051/0004-6361:20065375},
archivePrefix = {arXiv},
       eprint = {astro-ph/0607114},
 primaryClass = {astro-ph},
       adsurl = {https://ui.adsabs.harvard.edu/abs/2006A\&A...456..215F}
}

@ARTICLE{Bacmann2016,
       author = {{Bacmann}, A. and {Garc{\'\i}a-Garc{\'\i}a}, E. and {Faure}, A.},
        title = "{Detection of protonated formaldehyde in the prestellar core L1689B}",
      journal = {Astron. Astrophys.},
         year = 2016,
        month = apr,
       volume = {588},
          eid = {L8},
        pages = {L8},
          doi = {10.1051/0004-6361/201628280},
archivePrefix = {arXiv},
       eprint = {1603.00477},
 primaryClass = {astro-ph.GA},
       adsurl = {https://ui.adsabs.harvard.edu/abs/2016A\&A...588L...8B}
}

@article{Jimenez2025,
author = {Jim{\'e}nez-Redondo, Miguel and Sipilä, Olli and Dahl, Robin and Caselli, Paola and Jusko, Pavol},
title = {Cl+ and HCl+ in Reaction with H2 and Isotopologues: A Glance into H Abstraction and Indirect Exchange at Astrophysical Conditions},
journal = {ACS Earth and Space Chem.},
volume = {9},
number = {3},
pages = {782-788},
year = {2025},
doi = {10.1021/acsearthspacechem.4c00414},
URL ={https://doi.org/10.1021/acsearthspacechem.4c00414},
eprint ={https://doi.org/10.1021/acsearthspacechem.4c00414}
}

@article{oberg2023,
   author = "Öberg, Karin I. and Facchini, Stefano and Anderson, Dana E.",
   title = "Protoplanetary Disk Chemistry", 
   journal= "Annu. Rev. Astron. Astrophys.",
   year = "2023",
   volume = "61",
   number = "Volume 61, 2023",
   pages = "287-328",
   doi = "https://doi.org/10.1146/annurev-astro-022823-040820",
   url = "https://www.annualreviews.org/content/journals/10.1146/annurev-astro-022823-040820",
   publisher = "Annual Reviews",
   issn = "1545-4282",
   type = "Journal Article",
  }

@ARTICLE{Boogert2015,
       author = {{Boogert}, A.~C. Adwin and {Gerakines}, Perry A. and {Whittet}, Douglas C.~B.},
        title = "{Observations of the icy universe.}",
      journal = {Annu. Rev. Astron. Astrophys.},
         year = 2015,
        month = aug,
       volume = {53},
        pages = {541-581},
          doi = {10.1146/annurev-astro-082214-122348},
archivePrefix = {arXiv},
       eprint = {1501.05317},
 primaryClass = {astro-ph.GA},
       adsurl = {https://ui.adsabs.harvard.edu/abs/2015ARA&A..53..541B}
}

@ARTICLE{Harju2025,
       author = {{Harju}, J. and {Caselli}, P. and {Sipil{\"a}}, O. and {Spezzano}, S. and {Belloche}, A. and {Bizzocchi}, L. and {Pineda}, J.~E. and {Redaelli}, E. and {Wyrowski}, F.},
        title = "{Statistical nuclear spin ratios of deuterated ammonia in the pre-stellar core L1544}",
      journal = {Astron. Astrophys.},
         year = 2025,
        month = aug,
       volume = {700},
          eid = {A141},
        pages = {A141},
          doi = {10.1051/0004-6361/202555336},
archivePrefix = {arXiv},
       eprint = {2507.04825},
 primaryClass = {astro-ph.GA},
       adsurl = {https://ui.adsabs.harvard.edu/abs/2025A&A...700A.141H}
}

@ARTICLE{Wakelam2024,
       author = {{Wakelam}, V. and {Gratier}, P. and {Loison}, J.-C. and {Hickson}, K.~M. and {Penguen}, J. and {Mechineau}, A.},
        title = "{The 2024 KIDA network for interstellar chemistry}",
      journal = {Astron. Astrophys.},
         year = 2024,
        month = sep,
       volume = {689},
          eid = {A63},
        pages = {A63},
          doi = {10.1051/0004-6361/202450606},
archivePrefix = {arXiv},
       eprint = {2407.15958},
 primaryClass = {astro-ph.GA},
       adsurl = {https://ui.adsabs.harvard.edu/abs/2024A&A...689A..63W}
}

@ARTICLE{Ruane2017,
       author = {{Ruane}, G. and {Mawet}, D. and {Kastner}, J. and {Meshkat}, T. and {Bottom}, M. and {Femen{\'\i}a Castell{\'a}}, B. and {Absil}, O. and {Gomez Gonzalez}, C. and {Huby}, E. and {Zhu}, Z. and {Jensen-Clem}, R. and {Choquet}, {\'E}. and {Serabyn}, E.},
        title = "{Deep Imaging Search for Planets Forming in the TW Hya Protoplanetary Disk with the Keck/NIRC2 Vortex Coronagraph}",
      journal = {Astron. J.},
         year = 2017,
        month = aug,
       volume = {154},
       number = {2},
          eid = {73},
        pages = {73},
          doi = {10.3847/1538-3881/aa7b81},
archivePrefix = {arXiv},
       eprint = {1706.07489},
 primaryClass = {astro-ph.EP},
       adsurl = {https://ui.adsabs.harvard.edu/abs/2017AJ....154...73R}
}



\section*{Supplementary material}

# Physical structure of the disk models

Figure S1 shows the vertical temperature and density profiles in the three models used in this work, and are compared to the model of Lee et al.[S1] which is taken as a reference.

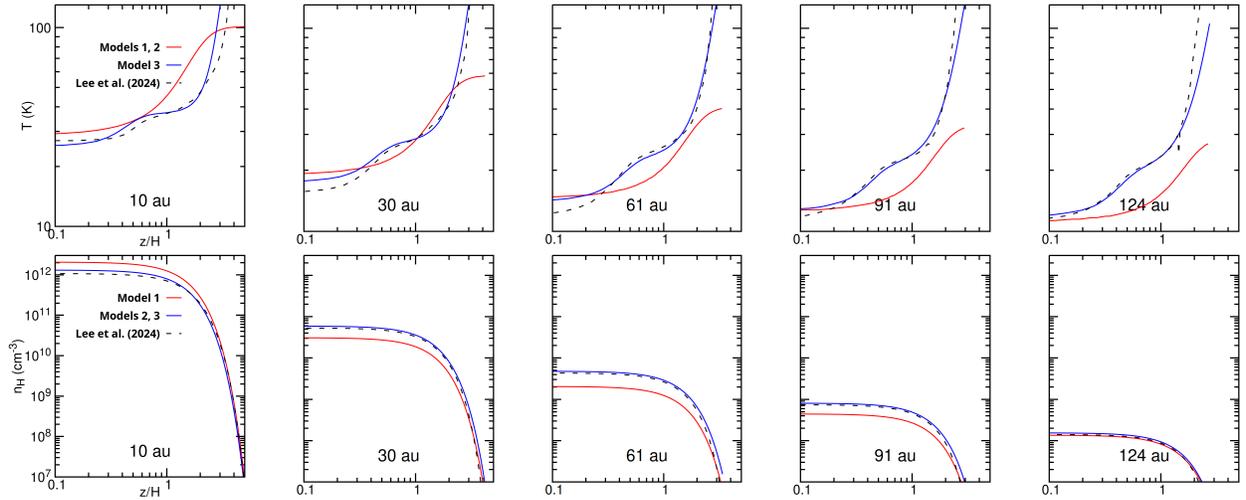

Figure S1: Comparison of the three parametric models (see Table 2 of the main paper) for the gas temperature (top) and H nuclei density (bottom). The vertical profiles are shown as function of $z/H$ for several representative radii indicated in each panel.

# Analytical ortho-to-para ratio of H$_2$CO

We derive below the analytical ortho-to-para ratio (OPR) of formaldehyde following its formation via the reaction of CH$_3$ with O($^3P$) atoms. Since the nuclear-spin thermalization of CH$_3^+$ by H$_2$ collisions should drive the OPR of CH$_3^+$ toward thermal equilibrium above a critical temperature $T_{\rm crit} \sim 15-20$ K (see Ref. 1), we will assume OPR(CH$_3^+$)=1 in the following.

Nuclear-spin branching ratios for the slow radiative association between CH$_3^+$ and H$_2$

can be derived using the approach of Quack[S2]:

$$pCH_3^+ + pH_2 \rightarrow \begin{cases} pCH_5^+ + h\nu & 0 \\ mCH_5^+ + h\nu & 1 \\ oCH_5^+ + h\nu & 0 \end{cases} \quad (1)$$

$$pCH_3^+ + oH_2 \rightarrow \begin{cases} pCH_5^+ + h\nu & 0 \\ mCH_5^+ + h\nu & \frac{1}{3} \\ oCH_5^+ + h\nu & \frac{2}{3} \end{cases} \quad (2)$$

$$oCH_3^+ + pH_2 \rightarrow \begin{cases} pCH_5^+ + h\nu & 0 \\ mCH_5^+ + h\nu & 0 \\ oCH_5^+ + h\nu & 1 \end{cases} \quad (3)$$

$$oCH_3^+ + oH_2 \rightarrow \begin{cases} pCH_5^+ + h\nu & \frac{1}{2} \\ mCH_5^+ + h\nu & \frac{1}{6} \\ oCH_5^+ + h\nu & \frac{1}{3} \end{cases} \quad (4)$$

From the above equations, we derive the OPR, meta-to-para ratio (MPR) and ortho-to-meta ratio (OMR) of $CH_5^+$:

$$\mathrm{OPR}(CH_5^+) = \frac{2\,\mathrm{OPR}(H_2) + 2}{\mathrm{OPR}(H_2)} \quad (5)$$

$$\mathrm{MPR}(CH_5^+) = \frac{2 + \mathrm{OPR}(H_2)}{\mathrm{OPR}(H_2)} \quad (6)$$

$$\mathrm{OMR}(CH_5^+) = \frac{2\,\mathrm{OPR}(H_2) + 2}{2 + \mathrm{OPR}(H_2)} \quad (7)$$

Now, by combining the above equations with the nuclear-spin branching ratio for the DR of

$CH_5^+$:

$$pCH_5^+ + e^- \rightarrow \begin{cases} pCH_3 + pH_2 & 0 \\ pCH_3 + oH_2 & 0 \\ oCH_3 + pH_2 & 0 \\ oCH_3 + oH_2 & 1 \end{cases} \quad (8)$$

$$mCH_5^+ + e^- \rightarrow \begin{cases} pCH_3 + pH_2 & \frac{2}{5} \\ pCH_3 + oH_2 & \frac{2}{5} \\ oCH_3 + pH_2 & 0 \\ oCH_3 + oH_2 & \frac{1}{5} \end{cases} \quad (9)$$

$$oCH_5^+ + e^- \rightarrow \begin{cases} pCH_3 + pH_2 & 0 \\ pCH_3 + oH_2 & \frac{1}{2} \\ oCH_3 + pH_2 & \frac{1}{4} \\ oCH_3 + oH_2 & \frac{1}{4} \end{cases} \quad (10)$$

we obtain:

$$\text{OPR}(CH_3) = \frac{7 + 11\text{OPR}(H_2)}{13 + 9\text{OPR}(H_2)} \quad (11)$$

By combining Eq. 11 with the branching ratio for the formation of $H_2CO$:

$$pCH_3 + O \rightarrow \begin{cases} pH_2CO + H & \frac{1}{2} \\ oH_2CO + H & \frac{1}{2} \end{cases} \quad (12)$$

$$oCH_3 + O \rightarrow \begin{cases} pH_2CO + H & 0 \\ oH_2CO + H & 1 \end{cases} \quad (13)$$

we finally obtain:

$$\text{OPR}(H_2CO) = 1 + 2\text{OPR}(CH_3) = \frac{27 + 31\text{OPR}(H_2)}{13 + 9\text{OPR}(H_2)} \quad (14)$$

We note that if OPR($H_2$)=3, the above equations predict that OPR($CH_3$)=1 and OPR($H_2CO$)=3, as expected in the statistical limit. On the other hand, in a *para*-rich $H_2$ gas (OPR($H_2$)$\ll$ 1), we predict OPR($CH_3$)=7/13 $\sim$ 0.54 and OPR($H_2CO$)=27/13 $\sim$ 2.1.

## OPRs of $H_2$, $CH_3$ and $H_2CO$

As an application of the above analytical treatment, we show in Fig. S2, the OPRs of $H_2$, $CH_3$ and $H_2CO$ as function of kinetic temperature for typical density and ionization conditions in the TW Hya protoplanetary disk: $n_H = 10^8 cm^{-3}$ and $\zeta = 5 \times 10^{-19}$ s$^{-1}$, and with the fixed depletion factors $\delta_C = 30$ and $\delta_N = 1.5$. The chemistry was first solved at steady-state with the UGAN network for kinetic temperatures in the range $5 - 50$ K and the above density and ionization rate. Equations (11) and (14) were then employed to compute the analytical OPRs of $CH_3$ and $H_2CO$, respectively, as function of the UGAN OPR of $H_2$ plotted in Fig. 1 (left panel). We can first observe that OPR($H_2$) is suprathermal below $\sim$15 K and closely follows thermal equilibrium at higher temperatures, as expected (see Faure et al.[S3]). In contrast, the UGAN OPRs of $CH_3$ and $H_2CO$ deviate significantly from thermal values, with a relatively weak temperature dependence. In particular, the OPR of $H_2CO$ never goes below 1.8, is suprathermal below 11 K, and increases towards the statistical value beyond 50 K. The good agreement between the UGAN results and the analytic calculations clearly demonstrates that the OPR of $H_2CO$ is governed by the OPR of $H_2$, via the formation of $CH_5^+$ (Eqs. 1-4) and $CH_3$ (Eqs. 8-10). We note, however, that the analytical OPRs slightly departs from the UGAN values. This can be explained by the actual temperature dependence of the OPR of $CH_3^+$, which drops below unity at temperatures lower than $\sim$ 25 K, and also by the secondary formation reactions of $CH_5^+$ and $H_2CO$ via $CH_4 + H_3^+$ and $H_2COH^+ + e^-$, respectively, which are neglected in our analytical treatment.

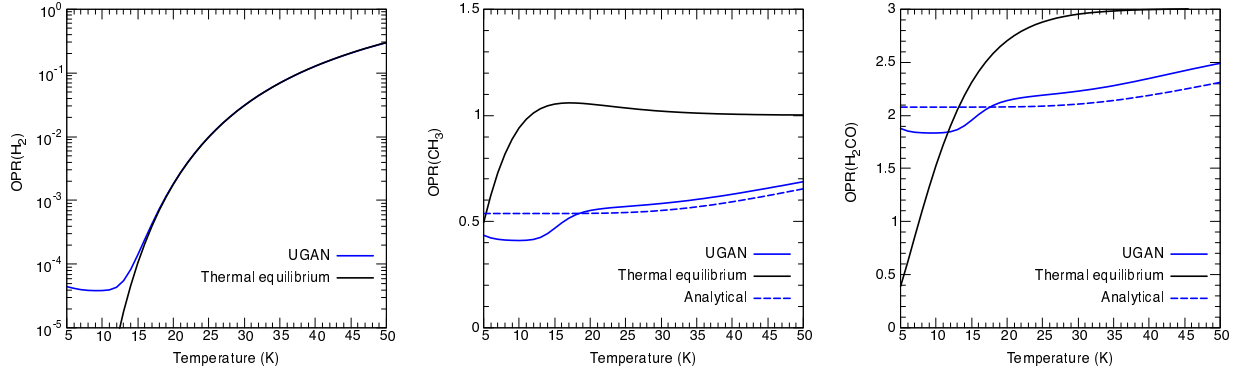

Figure S2: OPRs of $H_2$, $CH_3$ and $H_2CO$ as function of kinetic temperature for $n_H = 10^8 \text{cm}^{-3}$ and $\zeta = 5 \times 10^{-19}$ s$^{-1}$, with depletion factors $\delta_C = 30$ and $\delta_N = 1.5$. The `UGAN` results are compared to thermal equilibrum and, except for $H_2$, to the analytical model. See text for details.

# References

(S1) Lee, S.; Nomura, H.; Furuya, K. Carbon Isotope Chemistry in Protoplanetary Disks: Effects of C/O Ratios. *Astrophys. J.* **2024**, *969*, 41.

(S2) Quack, M. Detailed symmetry selection rules for reactive collisions. *Mol. Phys.* **1977**, *34*, 477–504.

(S3) Faure, A.; Hily-Blant, P.; Le Gal, R.; Rist, C.; Pineau des Forêts, G. Ortho-Para Selection Rules in the Gas-phase Chemistry of Interstellar Ammonia. *Astrophys. J.* **2013**, *770*, L2.



\end{document}